\documentclass[%
 reprint,onecolumn,notitlepage,
superscriptaddress,
 amsmath,amssymb,
 aps,
]{revtex4-1}
\usepackage{enumitem}
\usepackage{graphicx}
\usepackage{dcolumn}
\usepackage{bm}
\usepackage[colorinlistoftodos]{todonotes}
\usepackage{dsfont}
\usepackage{mathtools}
\usepackage{mathrsfs}
\usepackage{amsmath}
\usepackage{braket}
\usepackage{amsthm}
\DeclareMathOperator{\tr}{tr}
\newcommand{\norm}[1]{\left\lVert#1\right\rVert}
\DeclareMathOperator{\sech}{sech}
\DeclareMathOperator{\arctanh}{arctanh}

\usepackage{algorithm}
\usepackage{algpseudocode}
\algrenewcommand\algorithmicrequire{\textbf{Input:}}
\algrenewcommand\algorithmicensure{\textbf{Output:}}

\newtheorem{theorem}{Theorem}
\newtheorem{lemma}[theorem]{Lemma}

\theoremstyle{definition}
\newtheorem{definition}[theorem]{Definition}

\def\Item$#1${\item $\displaystyle#1$
   \hfill\refstepcounter{equation}(\theequation)}

\begin{document}

\title{Efficient verification
    of bosonic quantum channels via benchmarking}

\author{Ya-Dong Wu}
\affiliation{Institute for Quantum Science and Technology, University of Calgary, Alberta T2N 1N4,
Canada}
\author{Barry C. Sanders}
\affiliation{Institute for Quantum Science and Technology, University of Calgary, Alberta T2N 1N4,
Canada}
\affiliation{Program in Quantum Information Science, Canadian Institute for Advanced Research, Toronto, Ontario M5G 1M1, Canada} 
\affiliation{Shanghai Branch, National Laboratory for Physical Sciences at Microscale, University of Science and Technology of China, Shanghai
201315, People's Republic of China}

\date{\today}
\begin{abstract}
We aim to devise feasible,
efficient verification schemes for bosonic channels.
To this end,
we construct an average-fidelity witness
that yields a tight lower bound
for average fidelity
plus a general framework for verifying optimal quantum channels. 
For both multi-mode unitary Gaussian channels and single-mode amplification channels,
we present experimentally feasible average-fidelity witnesses and reliable verification schemes,
for which sample complexity scales polynomially with respect to all channel specification parameters.
Our verification scheme provides an approach to benchmark the performance of bosonic channels on a set of Gaussian-distributed coherent states by employing only two-mode squeezed vacuum states and local homodyne detections. 
Our results demonstrate how to perform feasible tests of quantum components designed for continuous-variable quantum information processing.
\end{abstract}
\maketitle


\section{\label{sec:intro}Introduction}
Progress in optical quantum computing~\cite{masada2015continuous,andersen2015hybrid,takeda2017universal} demands efficient schemes
to verify performance of optical quantum processes,
which would serve as components and devices for the quantum system.
Characterization by quantum process tomography~\cite{chuang1997prescription,poyatos1997complete,d2001quantum,altepeter2003ancilla,o2004quantum,lobino2008complete,rahimi2011quantum}
could serve as a means for gathering sufficient assessment data to be used for verification,
but, unfortunately,
quantum process tomography is inefficient:
the sampling overhead scales exponentially with system size,
with system size being logarithmic in Hilbert space dimension
corresponding to how much quantum information (e.g., number of qubits)
required to describe the system.
Direct fidelity estimation~\cite{flammia2011direct,da2011practical}
provides a way to partially characterize quantum channels with less overhead, but its adaption to bosonic channels requires measuring the Wigner function of output states at each phase-space point, and hence is not feasible due to the non-compactness of phase space.
Randomized benchmark~\cite{magesan2011scalable,magesan2012characterizing,wallman2014randomized,proctor2017randomized} provides a scalable method to evaluate the average performance of Clifford gates, however, its adaption to bosonic channels is not readily obtained because Gaussian unitary operations, as continuous-variable analog of Clifford gates, do not form an exact unitary $2$-design~\cite{zhuang2019scrambling}.
Our aim is to devise efficient and experimentally feasible verification schemes for bosonic channels.
 
Quantum-state verification is widely studied ~\cite{aolita2015reliable,hangleiter2017direct,takeuchi2018verification,gluza2018fidelity,pallister2018optimal,pallister2018optimal,zhu2018efficient}.
Reliable and efficient verification schemes~\cite{aolita2015reliable} for both bosonic Gaussian pure states and pure states generated by photon-number state inputs, linear optical interferometers and photon number detections
has been generalized to non-Gaussian cubic phase states~\cite{liu2018client}. 
These verification approaches have been adapted 
to benchmarking continuous-variable (CV) quantum gates~\cite{farias2018average}. On the other hand, a series of quantum-process benchmark approaches for bosonic channels have been explored~\cite{chiribella2013optimal,chiribella2014quantum,yang2014certifying,bai2018test}.
An alternative approach
benchmarks the average fidelity of bosonic quantum processes over all coherent states
by preparing a two-mode squeezed vacuum state and measuring a single observable~\cite{bai2018test}.

An experimentally appealing adaptation~\cite{farias2018average}
of recent verification schemes~\cite{aolita2015reliable,liu2018client} only estimates average fidelity over a finite-dimensional subspace chosen by selecting a finite set of coherent states.
This subspace selection cannot assess quantum-channel performance
over the entire infinite-dimensional Hilbert space~$\mathscr{H}$. 
In contrast,
the alternative scheme~\cite{bai2018test}
is challenged by  experimental limitations: 
online
squeezing,
which squeezes \emph{any} state known or unknown~\cite{yoshikawa2007demonstration,miwa2014exploring},
and quantum memories~\cite{lvovsky2009optical,saglamyurek2011broadband}. 
Here we combine the favourable features of the state verification approach~\cite{aolita2015reliable} and the unified quantum-benchmark approach~\cite{bai2018test} to develop our verification schemes for bosonic channels.

We formulate quantum-channel verification 
as an adversarial game between
a technology-limited verifier
and an untrusted, powerful prover
who has significant but bounded quantum technology.  
Our average-fidelity witness
issues a certificate that contains
a tight lower bound of the average fidelity of the quantum channel.
We develop a general framework for verification of optimal quantum channels,
and,
as examples of this framework,
we present reliable and experimentally feasible verification schemes for both multi-mode Gaussian unitary channels and single-mode amplification channels.
Both schemes can be implemented by preparing two-mode squeezed vacuum states and applying local homodyne detections, and the sample complexities for both two schemes scale polynomially with all channel-specification parameters.
Thus, our results provide experimentally feasible tests of quantum components in bosonic quantum systems.

Our paper is organized as follows. Section~\ref{sec:back} reviews the background related to Gaussian quantum information, verification of Gaussian pure states and benchmark of quantum channels. Section~\ref{sec:def} provides the mathematical definitions of verification of quantum channels and average-fidelity witness.
In Sec.~\ref{sec:result}, we introduce the verification scheme of multi-mode Gaussian unitary channels and single-mode amplification channels. Sections~\ref{sec:dis} and~\ref{sec:con} are the discussion and  conclusion sections, respectively.

\section{\label{sec:back}Background}
In this section, we first briefly review CV quantum information. Second, we present the mathematical definitions of quantum-state verification and of a fidelity witness,
and discuss verification protocols for multi-mode Gaussian pure states~\cite{aolita2015reliable}.
Third is a review of the unified benchmark approach~\cite{bai2018test}
by preparing one single input state and measuring one single observable. 

\subsection{Gaussian quantum information in a nutshell}

This subsection begins with pertinent basic concepts of CV quantum information~\cite{weedbrook2012gaussian,serafini2017quantum}.
We discuss the important concepts on Gaussian quantum information, necessary for the verification protocols in Sec.~\ref{sec:result}, including Gaussian states, multi-mode Gaussian unitary operations, and homodyne measurements. 
In bosonic systems, CV quantum information is encoded in the Fock space
$\mathscr{H}^{\otimes N}$,
where $\mathscr{H}$ is a single-mode Fock space, spanned by Fock number states
$\{\ket{n}\}_{n=0}^\infty$, and $N$ denotes the number of modes.
For each mode~$j$, we denote the position operator and momentum operator by~$\hat{q}_j$ and~$\hat{p}_j$, respectively.
The annihilation and creation operators are 
\begin{equation}\label{creationannihilation}
     \hat{a}_j\coloneqq\frac{\hat{q}_j+\mathrm{i}\hat{p}_j}{\sqrt{2}}, \;
      \hat{a}^\dagger_j\coloneqq\frac{\hat{q}_j-\mathrm{i}\hat{p}_j}{\sqrt{2}},\;
     \left[\hat{a}_j, \hat{a}^\dagger_k\right]=\delta_{jk}
\end{equation}
with the commutator describing the bosonic algebra.

Each density operator on~$\mathscr{H}$ is a trace-class operator.
Given an observable~$O$, its mean value is
\begin{equation}
    \braket{O}_{\rho}\coloneqq\tr(O\rho)=\sum_{n=0}^\infty \bra{n}O\ket{n}.
\end{equation}
To make $\tr(O\rho)$ well defined for any~$\rho$ on $\mathscr{H}$, either $O$ is bounded
or a sequence of bounded self-adjoint operator~$O^{(n)}$ exists such that $\forall \ket{\psi}\in \mathscr H$~\cite{hall2013quantum}
\begin{equation}\label{limitation}
    \norm{O^{(n)}\ket{\psi}-O\ket{\psi}}\to 0, \text{ as } n\to \infty,
\end{equation}
where $\norm{\cdot}$ is the Euclidean norm on~$\mathscr H$. For example, 
although the number operator $\hat{n}$ is not bounded, due to the finite energy restriction, a sequence of operators
\begin{equation}
    \left\{ \sum_{n=0}^m n\ket{n}\bra{n} \right\}_{m=0}^\infty
\end{equation}
approaches the limit in~(\ref{limitation}), so the mean photon number $\tr(\hat{n}\rho)$ is always well defined.

An $N$-mode density operator~$\rho$ is a Gaussian state if its characteristic function 
\begin{equation}
\label{eq:chi}
\chi(\bm{\xi})=\tr(\rho D(\bm{\xi})),\;
D(\bm{\xi})\coloneqq\mathrm{e}^{\mathrm{i}\hat{\bm{x}}^\mathrm{T}\bm{\Omega} \bm{\xi}},\;
\bm{\Omega}\coloneqq\bigoplus^N \begin{bmatrix}
0&1\\-1&0
\end{bmatrix},\;
\hat{\bm{x}}\coloneqq(\hat{q}_1, \hat{p}_1, \dots, \hat{q}_N, \hat{p}_N)^\top,\;
\bm{\xi} \in\mathbb{R}^{2N}
\end{equation}
is a Gaussian distribution,
where~${}^\top$ denotes transpose.
Any $N$-mode Gaussian state can be characterized by the first two cumulants of the conjugated quadrature operators,
i.e., mean vector and covariance matrix 
\begin{equation}
\bar{\bm{x}}\coloneqq\langle \hat{\bm{x}}\rangle,\;
V_{ij}\coloneqq\frac{1}{2}\braket{\{ \hat{\bm{x}}_i-\bar{\bm{x}}_i, \hat{\bm{x}}_j-\bar{\bm{x}}_j\} },
\end{equation}
where
\begin{equation}
    \{A, B\}\coloneqq AB+BA
\end{equation}
is the anti-commutator.

A thermal state at temperature~$T$
is a Gaussian state with density operator on Fock basis
\begin{equation}
    \rho_{T}(\bar{n}_T)
        =\sum_{n=0}^\infty \frac{\bar{n}_T^n}{(\bar{n}_T+1)^{n+1}}\ket{n}\bra{n},\;
    \bar{n}_T\coloneqq\frac{1}{\mathrm{e}^{\frac{\hbar\omega}{k_B T}}-1}
\end{equation}
where $\bar{n}_T$ is the mean photon number, $\omega$ is the frequency for this mode, and $k_B$ is Boltzmann's constant.
The density operator of a thermal state
can be represented as a function of annihilation and creation operators~\cite{collett1988exact,fan2003operator}, 
\begin{equation}\label{operatorRepresentationThermal}
    \rho_{T}(\bar{n}_T) 
    =\frac{1}{\bar{n}_T+1}\sum_{n=0}^{\infty}\frac{(-1)^n \hat{a}^{\dagger n}\hat{a}^n}{n!(\bar{n}_T+1)^n}.
\end{equation}
The purification of~$\rho_T(\bar{n}_T)$ is a two-mode squeezed vacuum state
\begin{equation}
\label{twoModeSqueezedVacuum}
    \ket{r}_{\mathrm{TMSV}}
    =\operatorname{csch}r
    \sum_{n=0}^\infty \tanh^n r \ket{n}\ket{n},\;
r=\arctanh\left(\sqrt{\frac{\bar{n}_T}{\bar{n}_T+1}}\right),
\end{equation}
for~$r$ the squeezing parameter. 

Gaussian unitary operations
(that is, unitary representations of Gaussian maps)
are unitary operations that map Gaussians quantum states to Gaussian quantum states.
Gaussian-preserving unitary operations form the semidirect product group~\cite{bartlett2002efficient}
\begin{equation}
\label{eq:HWSp}
    \mathrm{HW}(N)\rtimes \mathrm{Sp}(2N,\mathbb{R})=\{U_{\bm{S}, \bm{d}}; \bm{S}\in\mathrm{Sp}(2N,\mathbb{R}), \bm{d}\in \mathbb{R}^{2N} \}
\end{equation}
for $\mathrm{HW}(N)$ the Heisenberg-Weyl group comprising displacement operations on $N$-mode phase space
and $\mathrm{Sp}(2N,\mathbb{R})$ the real symplectic group comprising squeezers and linear optical interferometers. 
The spectral norm of~$\bm{S}$, denoted by $\norm{\bm{S}}_{\infty}$, equals~$\mathrm{e}^{r_\mathrm{max}}$, where $r_\mathrm{max}$ is the maximal single-mode squeezing parameter
in~$U_{\bm{S}, \bm{d}}$. 
Any multi-mode Gaussian unitary operation~$U_{\bm{S}, \bm{d}}$ yields
an affine mapping on phase space
\begin{equation}
\label{GaussianUnitary}
    \hat{\bm{x}}\to \bm{S}\hat{\bm{x}}+\bm{d}.
\end{equation}
 Under the Gaussian unitary operation $U_{\bm{S},\bm{d}}$, the mean values and covariance matrix of a Gaussian state are transformed to
\begin{equation}
    \bar{\bm{x}}\to\bm{S}\bar{\bm{x}}+\bm{d},\;
    \bm{V}\to \bm{S}\bm{V}\bm{S}^\top.
\end{equation}
The Gaussian unitary operation
\begin{equation}
    S_{\theta}\coloneqq\exp{\frac{\theta}{2}
    \left(\hat{a}_1\hat{a}_2+\hat{a}^\dagger_1\hat{a}^\dagger_2\right)}
\end{equation} 
with phase-space transformation 
\begin{equation}\label{two-modeSqueezingTransformation}
    \hat{\bm{x}}\to \bm{S}_\theta\hat{\bm{x}}, \;
    \bm{S}_\theta=\begin{bmatrix}
       \cosh\theta \bm{\mathds{1}} & \sinh\theta \bm{Z} \\
       \sinh\theta \bm{Z} & \cosh\theta \bm{\mathds{1}}
    \end{bmatrix},\;
    \bm{\mathds{1}}\coloneqq\begin{bmatrix} 
1&0\\
0&1
\end{bmatrix},\;
\bm{Z}\coloneqq\begin{bmatrix}
 1&0\\
 0 & -1
\end{bmatrix}
\end{equation}
is a two-mode squeezing operation. Online squeezing, in experiments, is the squeezing of an arbitrary, possibly unknown quantum state~\cite{weedbrook2012gaussian}.

Single-mode homodyne detection regarding quadrature operator
\begin{equation}
   \hat{x}_{\phi}= \cos \phi \hat{q}+
\sin\phi \hat{p}, \; \phi\in [0, \pi),
\end{equation}
corresponds to a positive operator-valued measurement (POVM)
\begin{equation}
    \{\ket{x}_{\phi}\bra{x}\mathrm{d}x; x\in\mathbb{R}\},
\end{equation} where $\ket{x}_{\phi}$ is an eigenstate of quadrature operator~$\hat{x}_{\phi}$
with eigenvalue~$x\in \mathbb{R}$, but not within~$\mathscr H$~\cite{de2005role}.
The probability of measurement outcome~$x$ is
\begin{equation}
  P_{\phi}(x)=\prescript{}{\phi}{\bra{x}} \rho\ket{x}_{\phi}.
\end{equation}
Experimentally, homodyne detection is accomplished by combining a signal mode with a local oscillator by a balanced beam splitter and
detecting the difference of photon numbers between the two output modes. 
Homodyne detection can be used for the purpose of
quantum tomography~\cite{lvovsky2009continuous}.

In this subsection, we have reviewed Gaussian states, multi-mode Gaussian unitary operations, as well as homodyne measurements. In the next subsection, we explain how to verify a Gaussian pure state. 

\subsection{Verification of pure states}
This subsection begins by the definition of quantum-state verification. Then we review the mathematical definition of fidelity witness~\cite{gluza2018fidelity}. Finally, we discuss the fidelity witness for Gaussian pure states and the verification protocol for Gaussian pure states~\cite{aolita2015reliable,farias2018average}. 

Verification is the process of determining whether an implementation properly satisfies design specifications~\cite{oberkampf2010verification}. 
Verification, along with validation that determines whether an implementation is qualified to accomplish a certain task, is important for assessing the credibility of a product or a system.
Here quantum-state verification~\cite{aolita2015reliable,hangleiter2017direct,takeuchi2018verification,gluza2018fidelity,pallister2018optimal,pallister2018optimal,zhu2018efficient} aims to check whether an implementation of certain quantum state meets the specifications of a target quantum state or not. 
While ref.~\cite{aolita2015reliable,hangleiter2017direct,gluza2018fidelity} use ``certification" to refer to the process of verification,
in this paper, we use the phrase ``quantum-state verification" rather than ``certification".

There is a technology-limited verifier and an untrusted, powerful prover with significant but bounded quantum technology.
The verifier provides the prover with the classical description of a pure state $\rho_\text{t}$, and the prover
sends independent and identical copies of quantum state~$\rho_{\mathrm{p}}$ to the verifier.
Then by measurements, the verifier decides whether to accept~$\rho_{\mathrm{p}}$ as a certified preparation of~$\rho_\text{t}$ or reject it. The figure of merit for state verification is fidelity 
\begin{equation}
    F(\rho_{\mathrm{p}}, \rho_\text{t})
        =\tr\left(\rho_{\mathrm{p}} \rho_\text{t} \right).
\end{equation}
Reminiscent of interactive proof systems~\cite{goldwasser1989knowledge,homer2011computability}, the completeness and soundness conditions of quantum-state verification are defined as follows.

\begin{definition}[\cite{aolita2015reliable}]
\label{definitionVerificationQuantumState}
With respect to threshold fidelity~$F_\text{t}<1$ and maximal failure probability~$0<\delta\le\frac{1}{2}$, the verifier's verification test should satisfy
\begin{enumerate}
  \item completeness: if~$\rho_{\mathrm{p}}=\rho_\text{t}$, the verifier accepts with probability at least~$1-\delta$;
  \item soundness: if~$F(\rho_{\mathrm{p}}, \rho_\text{t})\le F_\text{t}$, the verifier rejects with probability at least~$1-\delta$.
  \end{enumerate}
\end{definition}
\noindent As $\rho_\text{t}$ has zero measure in the topological space of density operators induced by fideity, to make the definition practically meaningful, the verifier should accept all states in a neighbourhood of~$\rho_\text{t}$ with probability at least~$1-\delta$. 

In the multi-qubit case, $F(\rho_{\mathrm{p}}, \rho_\text{t})$ can be estimated~\cite{flammia2011direct,da2011practical} by decomposing~$\rho_\text{t}$ into a linear combination of Pauli operators and measuring the overlap between~$\rho_{\mathrm{p}}$ and each Pauli operator. This idea gives rise to verification schemes for ground states of Hamiltonians and certain stabilizer states by measuring single-qubit Pauli operators~\cite{takeuchi2018verification}. Adapting this idea into infinite-dimensional system, $F(\rho_{\mathrm{p}}, \rho_\text{t})$ can be estimated by measuring the Wigner function of~$\rho_{\mathrm{p}}$ at different phase-space points~\cite{da2011practical}. Although experimentally viable~\cite{lvovsky2009continuous}, as the phase space is non-compact, this method cannot yield a reliable estimation of the fidelity of a CV state by using a finite number of copies.

To obtain an efficient verification scheme for Gaussian pure states, we introduce fidelity witness, which provides an economic way to detect~$F(\rho_{\mathrm{p}}, \rho_\text{t})$. Analogous to entanglement witness~\cite{terhal2000bell,horodecki2009quantum}, a fidelity witness distinguishes~$\rho_\text{t}$ from the whole set~$\{\rho_{\mathrm{p}}; F(\rho_{\mathrm{p}}, \rho_\text{t})\le F_\text{t}\}$ for any threshold fidelity~$F_\text{t}<1$.
 Here we present the mathematical definition of fidelity witness.
\begin{definition}[\cite{gluza2018fidelity}]
A self-adjoint operator~$W$ is a fidelity witness for~$\rho_\text{t}$ if
\begin{equation}
\omega(\rho_{\mathrm{p}})\coloneqq\operatorname{tr}\left(W\rho_{\mathrm{p}} \right)    
\end{equation}
satisfies
    \begin{enumerate}
        \Item $\omega(\rho_{\mathrm{p}})=1\Longleftrightarrow\rho_{\mathrm{p}}=\rho_\text{t};$
        \Item $\forall \rho_{\mathrm{p}}, \omega(\rho_{\mathrm{p}})\le F(\rho_{\mathrm{p}}, \rho_\text{t}).$
    \end{enumerate}
\end{definition}
\noindent We see that 
\begin{equation}
    \operatorname{tr}\left(W\rho_{\mathrm{p}} \right)> F_\text{t}
\end{equation} witnesses 
\begin{equation}
    F(\rho_\text{t}, \rho_{\mathrm{p}})> F_\text{t},
\end{equation} 
whereas
\begin{equation}
    \operatorname{tr}\left(W\rho_{\mathrm{p}} \right)\le F_\text{t}
\end{equation} does not imply any relation between $F(\rho_\text{t}, \rho_{\mathrm{p}})$ and $F_\text{t}$. 

Now we explain how to verify a Gaussian pure state by measuring a fidelity witness, which has been first studied~\cite{aolita2015reliable} and then summarized in the formalism of fidelity witness~\cite{farias2018average}.
For any Gaussian pure state
\begin{equation}\label{GaussianPureTarget}
    \rho_\text{t}=U_{\bm{S},\bm{d}}\ket{0}\bra{0}U_{\bm{S}, \bm{d}}^\dagger,
\end{equation}
the observable
\begin{equation}\label{GaussianPureFidelityWitness}
     \mathds{1}- U_{\bm{S},\bm{d}}\hat{n}U_{\bm{S}, \bm{d}}^\dagger
\end{equation}
is a fidelity witness,
such that
\begin{equation}
     \label{witnessGaussianPureState}
   F(\rho_\text{t}, \rho_{\mathrm{p}}) \ge 1-\left\langle U_{\bm{S},\bm{d}}\hat{n}U_{\bm{S}, \bm{d}}^\dagger\right\rangle_{\rho_{\mathrm{p}}},
\end{equation}
where equality is achieved iff $\rho_{\mathrm{p}}=\rho_\text{t}$.
The above mean value is a linear combination of single-mode expectation values and two-mode correlations~\cite{aolita2015reliable}
\begin{equation}\label{GaussianPureWitnessReformulation}
    \left\langle U_{\bm{S},\bm{d}}\hat{n}U_{\bm{S}, \bm{d}}^\dagger\right\rangle_{\rho_{\mathrm{p}}}=
    \frac{1}{2}\tr \left[\bm{S}^{-\top}\bm{S}^{-1} \left(\left\langle\hat{\bm{x}}^\top\hat{\bm{x}}\right\rangle_{\rho_{\mathrm{p}}}-2\bar{\bm{x}}_{\rho_{\mathrm{p}}} \bm{d}+\bm{d}^\top\bm{d}\right) \right]-\frac{N}{2},
\end{equation}
Thus, the right-hand side of inequality~(\ref{witnessGaussianPureState}) can be estimated by local homodyne detections on~$\rho_{\mathrm{p}}$. 

The verification protocol for Gaussian pure states~\cite{aolita2015reliable} is reformulated in Algorithm~\ref{alg:GaussianPureState}. 
This protocol requires $2mc_1+2\nu mc_2$ copies of~$\rho_{\mathrm{p}}$~\cite{aolita2015reliable}, where
        \begin{align}
         &c_1\in O\left(\frac{m^2 \norm{\bm{S}}_{\infty}^4 \norm{\bm{d}}^2 \sigma_1^2}{\epsilon^2 \ln (1/(1-\delta))}\right), \\
         &c_2\in O\left(\frac{m^3 \nu^2 \norm{\bm{S}}_{\infty}^4  \sigma_2^2}{\epsilon^2 \ln (1/(1-\delta))}\right),
     \end{align} $\nu=2 \min\{k^2, m\}$, and
     $k$ is the maximum number of input modes to which an output mode is coupled.
\begin{algorithm}[H]
\caption{Verification protocol for Gaussian pure states~\cite{aolita2015reliable}} \label{alg:GaussianPureState}
    \begin{algorithmic}[1]    
    \Require{\Statex 
    \begin{itemize}
        \item $\bm{S}$ \Comment{$\bm{S}\in \mathrm{Sp}(2m,\mathbb{R})$}
        \item $\bm{d}$ \Comment{$\bm{d}\in \mathbb{R}^{2m}$}
        \item $F_\text{t}$ \Comment{$0<F_\text{t}<1$ is threshold fidelity}
        \item $\delta$ \Comment{$0<\delta\le\frac{1}{2}$ is maximal failure probability}
        \item $\epsilon$ \Comment{$0<\epsilon< \frac{1-F_\text{t}}{2}$ is error bound}
        \item $k$ \Comment{$k\in \mathbb{N}^+$ is the maximum number of input modes to which an output mode is coupled.}
        \item  $\rho_{\mathrm{p}}$ \Comment{$2mc_1+2\nu mc_2$ copies of~$\rho_{\mathrm{p}}$}
     \item $\sigma_1$ \Comment{The upper bound of the variance of any $\hat{\bm{x}}_l$ on $\rho_{\mathrm{p}}$, where $1\le l\le 2m$.} 
     \item $\sigma_2$ \Comment{The upper bound of the variance of any $\frac{1}{2}\left(\hat{\bm{x}}_u \hat{\bm{x}}_v+\hat{\bm{x}}_v\hat{\bm{x}}_u\right)$, where $1\le u\le v\le 2m$.}
        \end{itemize}
    }

    \Ensure{\Statex
    \begin{itemize}
    \item $b$ \Comment{$b\in\{0, 1\}$, $0$ means reject and $1$ means accept.}
    \end{itemize}
    }

    \Procedure{VerificationofPureGaussianStates}{$\bm{S}$, $\bm{d}$, $F_\text{t}$, $\delta$, $\epsilon$, $k$, $\sigma_1$, $\sigma_2$, $\rho_{\mathrm{p}}$}
 \For{$l=1: 2m$} 
 \For{$i=1:c_1$}\Comment{To obtain an estimate~$\bar{\bm{x}}_{\rho_{\mathrm{p}}}^*$ of~$\bar{\bm{x}}_{\rho_{\mathrm{p}}}$.}
\State apply a single-shot homodyne detection for quadrature $\hat{\bm{x}}_l$ on one copy of~$\rho_{\mathrm{p}}$;
\EndFor
 \State $\left(\bar{\bm{x}}_{\rho_{\mathrm{p}}}^*\right)_l\leftarrow \frac{1}{c_1} \sum_{i=1}^{c_1} \chi^{\hat{\bm{x}}_l}_i$; 
 \Comment{$\chi^{\hat{\bm{x}}_l}_i$ is $i$th measurement outcome with respect to $\hat{\bm{x}}_l$.}
\For{$i=1:c_2$} \Comment{To estimate the diagonal elements in $\left\langle\hat{\bm{x}}^\top\hat{\bm{x}}\right\rangle_{\rho_{\mathrm{p}}}$.}
\State apply a single-shot homodyne detection for quadrature $\hat{\bm{x}}_l$ on one copy of~$\rho_{\mathrm{p}}$;
\EndFor
\State $\left(\left\langle\hat{\bm{x}}^\top\hat{\bm{x}}\right\rangle_{\rho_{\mathrm{p}}}^*\right)_{ll}\leftarrow
 \frac{1}{c_2} \sum_{i=1}^{c_2} \left(\chi^{\hat{\bm{x}}_l}_i\right)^2$;
 \EndFor 
 
     \algstore{GaussianStateAlgPart1}
  \end{algorithmic}
 \end{algorithm}
 \begin{algorithm}[H]
  \begin{algorithmic}[1]
  \algrestore{GaussianStateAlgPart1}
  
 \For{$ v=1: 2m$} \Comment{To  estimate the off-diagonal elements in~$\left\langle\hat{\bm{x}}^\top\hat{\bm{x}}\right\rangle_{\rho_{\mathrm{p}}}$}
 \For{$u=1:v-1$ and $(\bm{S}^{-\top}\bm{S}^{-1})_{u,v}\neq 0$}
 \If{$(u, v)\neq (2j-1, 2j)$}
 \For{$i=1:c_2$}
 \State apply two single-shot homodyne detections for quadratures $\hat{\bm{x}}_u$ and $\hat{\bm{x}}_v$ simultaneously on one copy of~$\rho_{\mathrm{p}}$;
 \EndFor
 \State $\left(\left\langle\hat{\bm{x}}^\top\hat{\bm{x}}\right\rangle_{\rho_{\mathrm{p}}}^*\right)_{vu}\leftarrow
 \frac{1}{c_2} \sum_{i=1}^{c_2}\chi^{\hat{\bm{x}}_u}_i\chi^{\hat{\bm{x}}_v}_i$;
 \Comment{$\chi^{\hat{\bm{x}}_u}_i$ and $\chi^{\hat{\bm{x}}_v}_i$ are $i$th measurement outcomes}
  \State $\left(\left\langle\hat{\bm{x}}^\top\hat{\bm{x}}\right\rangle_{\rho_{\mathrm{p}}}^*\right)_{uv}\leftarrow
\left(\left\langle\hat{\bm{x}}^\top\hat{\bm{x}}\right\rangle_{\rho_{\mathrm{p}}}^*\right)_{vu}$;
 \Else
 \For {$i=1:c_2$}
\State apply a single-shot homodyne detection for quadrature $\frac{1}{\sqrt{2}}\left(\hat{\bm{x}}_u+\hat{\bm{x}}_v\right)$ on one copy of~$\rho_{\mathrm{p}}$;
\EndFor
\State $\left(\left\langle\hat{\bm{x}}^\top\hat{\bm{x}}\right\rangle_{\rho_{\mathrm{p}}}^*\right)_{vu}\leftarrow
 \frac{1}{c_2} \sum_{i=1}^{c_2}\left(\chi^{\frac{1}{\sqrt{2}}\left(\hat{\bm{x}}_u+\hat{\bm{x}}_v\right)}_i \right)^2 -\frac{1}{2}\left(\bar{\bm{x}}_{\rho_{\mathrm{p}}}^*\right)_u^2 -\frac{1}{2}\left(\bar{\bm{x}}_{\rho_{\mathrm{p}}}^*\right)_v^2$;
 \Statex\Comment{$\chi^{\frac{1}{\sqrt{2}}\left(\hat{\bm{x}}_u+\hat{\bm{x}}_v\right)}_i$ is $i$th measurement outcome regarding $\frac{1}{\sqrt{2}}\left(\hat{\bm{x}}_u+\hat{\bm{x}}_v\right)$.}
 \State $\left(\left\langle\hat{\bm{x}}^\top\hat{\bm{x}}\right\rangle_{\rho_{\mathrm{p}}}^*\right)_{uv}\leftarrow \left(\left\langle\hat{\bm{x}}^\top\hat{\bm{x}}\right\rangle_{\rho_{\mathrm{p}}}^*\right)_{vu}$;
 \EndIf
 \EndFor
 \EndFor

\State  $\omega(\rho_{\mathrm{p}})^*\leftarrow 
 \frac{1}{2}\tr \left[\bm{S}^{-\top}\bm{S}^{-1} \left(\left\langle\hat{\bm{x}}^\top\hat{\bm{x}}\right\rangle_{\rho_{\mathrm{p}}}^*-2\bar{\bm{x}}_{\rho_{\mathrm{p}}}^* \bm{d}+\bm{d}^\top\bm{d}\right) \right]-\frac{N}{2}$;
 \Comment{Obtain an estimate~$\omega(\rho_{\mathrm{p}})^*$  of 
 \begin{equation}
     \omega(\rho_{\mathrm{p}})=1-\left\langle U_{\bm{S},\bm{d}}\hat{n}U_{\bm{S}, \bm{d}}^\dagger\right\rangle_{\rho_{\mathrm{p}}},
 \end{equation} }
 \If{$\omega(\rho_{\mathrm{p}})^*>F_\text{t}+\epsilon$}
 \State \Return $b=1$;
 \Else \State \Return $b=0$.
 \EndIf
  \EndProcedure
    \end{algorithmic}
\end{algorithm}
\noindent
This protocol is a reliable verification protocol satisfying the completeness and soundness conditions in Def.~\ref{definitionVerificationQuantumState}.
Furthermore, this protocol accepts any state close enough to~$\rho_\text{t}$~\cite{aolita2015reliable}.
For any~$\rho_\text{t}$ and~$\rho_\text{p}$, if
\begin{equation}
F(\rho_{\mathrm{p}}, \rho_\text{t})\ge F_\text{t}+\Delta,
\end{equation}
where 
$0<\Delta<1-F_{\text{t}}$ is a fidelity gap~\cite{aolita2015reliable} depending on both~$\rho_\text{t}$ and~$\rho_\text{p}$,
the verifier accepts~$\rho_{\mathrm{p}}$ with probability at least~$1-\delta$.
As 
\begin{equation}
    F_\text{t}+\Delta<1,
\end{equation}
the verifier, with a high probability, accepts any state in a neighbourhood of~$\rho_\text{t}$ in the topological space of density operators. 

In this subsection, we have explained how verification of quantum states can be cast into an adversarial game between a verifier and a prover. We have
reviewed the mathematical definitions of fidelity witness as well as the verification protocol for multi-mode Gaussian pure states.

\subsection{Benchmarking quantum channels}\label{subsec:benchmark}
This subsection reviews the general framework of quantum-process benchmarking~\cite{yang2014certifying}. 
After that, we explain how an arbitrary benchmark test can be reformulated into a canonical test that employs one input state and measures one observable~\cite{bai2018test}. 

Here quantum-process benchmarking refers to measuring the performance of an experimental quantum process using a specific figure of merit, such as average fidelity, resulting in a value
that is compared with theoretical values. Direct-fidelity estimation approach~\cite{flammia2011direct,da2011practical} can be used to benchmark multi-qubit quantum channels by preparing product states and measuring single-qubit Pauli operators. On the other hand, quantum randomized benchmarking provides an efficient way to estimate the average gate fidelity of multi-qubit Clifford gates.
However, neither of these methods are readily adapted to benchmarking bosonic  channels due to the finite-energy restriction~\cite{chiribella2014quantum,yang2014certifying,bai2018test,sharma2018characterizing,farias2018average}.

Now we introduce a general framework of quantum-process benchmarking in terms of a quantum-state transformation game~\cite{yang2014certifying}.
In order to measure the performance of a prover's quantum channel, denoted by~$\mathcal E$, a verifier
prepares a state~$\rho_x$ with probability~$p_x$ (in general, a probability measure), sends~$\rho_x$ through~$\mathcal E$, 
applies certain measurement on~$\mathcal E(\rho_x)$, and assign different scores to different measurement outcomes,
where~$x$ is a label. 
We use~$x$ to denote the set of labels, and the cardinality of~$X$ can 
either be finite or be countably infinite or even uncountable. 
The expected score~$s_{\mathcal E}$ quantifies the performance of channel~$\mathcal E$.

For average-fidelity-based benchmarking, the verifier's measurement is described by the POVM
\begin{equation}
\{\ket{\phi_x}\bra{\phi_x}, \mathds{1}-\ket{\phi_x}\bra{\phi_x}\}, \; \ket{\phi_x}\in \mathscr{H}.    
\end{equation}
 If the measurement outcome corresponds to $\ket{\phi_x}\bra{\phi_x}$,
then the verifier assigns score~$1$ to~$\mathcal E$; otherwise he assigns score~$0$.
 Then the expected score equals the average fidelity
\begin{equation}\label{averageFidelityBenchmark}
    s_{\mathcal E}=\bar{F}_{\mathcal E} \coloneqq\sum_{x\in X} p_x \bra{\phi_x} \mathcal E(\rho_x)\ket{\phi_x},
\end{equation}
where, if~$x$ is an uncountable set, $\sum$ must be replaced by~$\int$.

Now we discuss a benchmark test, proposed in~\cite{bai2018test}, which requires only one input state and measurements of one observable. Rather than sampling different inputs~$\rho_x$, any benchmark test can be reformulated into a new test that requires only the preparation of one input state $\sigma_{\mathrm{AR}}$ and the measurement of one observable~$O_{\mathrm{A}'\mathrm{R}}$ by adding a reference system~R, where A and A$'$ denote channel input and channel output, respectively.  The new test is equivalent to the original one, in the sense that,
for any CPTP map~$\mathcal E$,
the expected score 
\begin{equation}\label{equivalentTest}
    s_{\mathcal E}=\tr{\left[O_{\mathrm{A}'\mathrm{R}} \mathcal E\otimes \mathcal{I} (\sigma_{\mathrm{AR}})\right]},
\end{equation}
where~$\mathds{1}$ is the identity channel on reference~R.
$\sigma_{\mathrm{AR}}$ and $O_{\mathrm{A}'\mathrm{R}}$ in Eq.~(\ref{equivalentTest}) are not unique:
different combinations of input~$\sigma_{\mathrm{AR}}$ and observable~$O_{\mathrm{A}'\mathrm{R}}$ lead to equivalent tests iff they yield the same performance operator~\cite{bai2018test}, which is defined below.
\begin{definition}[\cite{bai2018test}]
For a benchmark test with input state~$\sigma_{\mathrm{AR}}$ and observable~$O_{\mathrm{A}'\mathrm{R}}$, the performance operator is
\begin{equation}
\label{performanceOperator}
    \Omega_{\mathrm{A'A}}\coloneqq\tr_{\mathrm{R}}\left[(O_{\mathrm{A}'\mathrm{R}}\otimes \mathds{1}_{\mathrm{A}})(\mathds{1}_{\mathrm{A}'}\otimes \sigma_{\mathrm{AR}})\right].
\end{equation}
This performance operator~(\ref{performanceOperator})
satisfies the condition that,
for any quantum channel~$\mathcal E$,
\begin{equation}
    s_{\mathcal E}=\tr(\Omega_{\mathrm{A'A}}C_{\mathcal E}),
\end{equation}
for~$C_{\mathcal E}$ the Jamio\l{}kowski operator
for~$\mathcal E$~\cite{jamiolkowski1972linear}.
\end{definition}
Here we present one of the main results in~\cite{bai2018test}. As the combination of~$\sigma_{\mathrm{AR}}$
and~$O_{\mathrm{A}'\mathrm{R}}$ is not unique,
an experimentally feasible input state~$\sigma_{\mathrm{AR}}$ is preferred. 
Any benchmark test of~$\mathcal E$ can be reformulated into a canonical test by preparing an entangled pure state~$\ket{\Psi}_{\mathrm{AR}}$, applying~$\mathcal E$ to system~A, and applying measurements on $\mathcal E\otimes \mathcal{I} (\ket{\Psi}\bra{\Psi}_{\mathrm{AR}})$ with the observable~\cite{bai2018test}
\begin{equation}\label{benchmarkObservable}
         O_{\mathrm{A}'\mathrm{R}}=\left(\mathds{1}_{\mathrm{A}'}\otimes\rho_{\mathrm{R}}^{-\frac{1}{2}}T_{\mathrm{AR}}^\dagger\right)\Omega_{\mathrm{A'A}}^{\top_{\mathrm{A}}}\left(\mathds{1}_{\mathrm{A}'}\otimes T_{\mathrm{AR}}\rho_{\mathrm{R}}^{-\frac{1}{2}} \right),
     \end{equation}
     where 
     \begin{equation}
         \rho_{\mathrm{R}}=\tr_{\mathrm{A}}(\ket{\Psi}\bra{\Psi}_{\mathrm{AR}}), \; \rho_{\mathrm{A}}=\tr_{\mathrm{R}}(\ket{\Psi}\bra{\Psi}_{\mathrm{AR}})
     \end{equation} 
      and $T_{\mathrm{AR}}$ is a partial isometry such that 
      \begin{equation}
     T_{\mathrm{AR}}^\dagger \rho_{\mathrm{A}} T_{\mathrm{AR}}=\rho_{\mathrm{R}}.
     \end{equation}
By plugging the performance operator for average-fidelity-based test
\begin{equation}\label{average-fidelity-based-performance-operator}
    \Omega_{\mathrm{A'A}}=\sum_{x\in X}p_x \ket{\phi_x}\bra{\phi_x}\otimes \rho_x
\end{equation}
into Eq.~(\ref{benchmarkObservable}),
we obtain the single observable to be measured, in order to estimate average fidelity.


We have briefly reviewed CV quantum information theory, especially Gaussian states and Gaussian unitary operations.
Furthermore, we have reviewed concepts concerning quantum-state verification and fidelity witness.
Our exposition has elucidated how a multi-mode Gaussian pure state can be verified by measuring a fidelity witness. 
We have also discussed quantum-process benchmark and the canonical benchmark test.

\section{\label{sec:def}Definitions and framework}

This section develops our general framework of verification of an optimal quantum channel. We introduce a new concept, called average-fidelity witness. We present our general protocol for quantum-channel verification and show this verification protocol satisfies completeness and soundness conditions.

Consider a state-transformation task
\begin{equation}
\rho_x\mapsto \ket{\phi_x}
\end{equation}
with an input ensemble
\begin{equation}\label{InputEnsemble}
    \{(p_x, \rho_x); x\in X\}
\end{equation}
as well as an output-target-state set 
\begin{equation}\label{OutputEnsemble}
    \{\ket{\phi_x}\bra{\phi_x}; x\in X\}.
\end{equation} 
Suppose at least one optimal quantum channel~$\mathcal E_{\mathrm{opt}}$ exists in the sense that~$\mathcal E_{\mathrm{opt}}$ achieves the maximal average fidelity
\begin{equation}
    \bar{F}_{\mathrm{max}}\coloneqq\sup_{\mathcal E} \sum_{x\in X} p_x \bra{\phi_x} \mathcal E(\rho_x)\ket{\phi_x} =\sum_{x\in X} p_x \bra{\phi_x} \mathcal E_{\mathrm{opt}}(\rho_x)\ket{\phi_x}.
\end{equation} 
In the finite-dimensional case, such an optimal quantum channel always exists~\cite{konig2009operational,chiribella2013optimal}.

There is a technology-limited verifier and an untrusted, powerful prover with significant but bounded quantum technology.
The verifier provides the prover with the classical description of the input ensemble~(\ref{InputEnsemble}) as well as the output-target-state set~(\ref{OutputEnsemble}), and the prover
sends independent and identical copies of quantum channels,~$\mathcal E_{\mathrm{p}}$, to the verifier. 
The verifier prepares input states and applies local measurements at outputs without any state-preparation and measurement (SPAM) errors, 
and then decides whether to accept~$\mathcal E_{\mathrm{p}}$ as an optimal quantum channel in terms of~$\bar{F}_{\mathcal E_{\mathrm{p}}}$, or reject it.
We define completeness and soundness requirements for verification of optimal quantum channels 
as follows.

\begin{definition}\label{def:quantum-channel-verification}
An optimal-quantum-channel verification, with respect to threshold average fidelity~$\bar{F}_\text{t}$
and maximal failure probability~$\delta$, 
satisfies
\begin{enumerate}
\item completeness: if~$\bar{F}_{\mathcal E_{\mathrm{p}}}=\bar{F}_{\mathrm{max}}$,
then the verifier accepts with probability no less than~$1-\delta$;
  \item soundness: if~$\bar{F}_{\mathcal E_{\mathrm{p}}}\le \bar{F}_\text{t}$, then the verifier rejects with probability no less than~$1-\delta$.
  \end{enumerate}
\end{definition} 
\noindent To guarantee quantum-channel verification makes sense in practice, the verifier should accept any quantum channel in a neighbourhoood of~$\mathcal E_{\mathrm{opt}}$ in the topolocal space of all CPTP maps induced by the average fidelity in Eq.~(\ref{averageFidelityBenchmark}).

 In order to verify whether~$\mathcal E_{\mathrm{p}}$ is optimal,
 one way is to follow the procedures of the canonical average-fidelity-based benchmark test in Subsec.~\ref{subsec:benchmark}.
In general, however,~$O_{\mathrm{A}'\mathrm{R}}$ in Eq.~(\ref{benchmarkObservable}) is not feasibly measured.
Here we 
define average-fidelity witness, which yields a tight lower bound of the average fidelity
and develop a quantum-channel verification protocol involving measurement of an average-fidelity witness.
\begin{definition}\label{channelPerformanceWitness}
An observable~$W_{\mathrm{A}'\mathrm{R}}$ is an average-fidelity witness for $\bar{F}_{\mathcal E}$ on the state~$\mathcal E\otimes \mathcal I\left(\ket{\Psi}\bra{\Psi}_{\mathrm{AR}}\right)$ if 
\begin{equation}
\omega(\mathcal E)\coloneqq\tr\left[W_{\mathrm{A}'\mathrm{R}} \mathcal E\otimes \mathcal{I} \left(\ket{\Psi}\bra{\Psi}_{\mathrm{AR}}\right)\right]    
\end{equation}
satisfies
\begin{enumerate}
    \Item $\omega(\mathcal E)=\bar{F}_{\mathcal E} \Longleftrightarrow \bar{F}_{\mathcal E}=\bar{F}_{\mathrm{max}};$  \label{averagewitnessRequirement1}
    \Item $\forall \mathcal E, \omega(\mathcal E)\le \bar{F}_{\mathcal E}.$ \label{averagewitnessRequirement2}
\end{enumerate}
\end{definition}
\noindent Analogous to the fidelity witness,
measuring the average-fidelity witness distinguishes the optimal quantum channels from all quantum channels, whose average fidelity is below the threshold.


The verification game between the verifier and the prover can also be interpreted by a query model: copies of quantum channel~$\mathcal E$ are obtained via queries from a black box to decide whether~$\mathcal E$ is optimal or not in terms of average fidelity. Given certain classical descriptions of input and target-output ensembles, the black box, each time, outputs one independent and identical copy of a quantum channel. The query complexity describes how many copies of~$\mathcal E$ are demanded from the black box, in order to have a reliable answer on whether $\mathcal E$ is optimal or not. As estimating the mean value of an average-fidelity witness is sampling the mean value of an unknown distribution, we use sampling complexities, instead of query complexities, from now on, to infer how the number of copies of~$\mathcal E$ scales with respect to the size of the classical description of input and target-output ensembles. 
We present our general framework of a verification protocol for optimal quantum channels in Algorithm~\ref{alg:generalprotocol}.

\begin{algorithm}[H]
    \begin{algorithmic}[1]   
    \caption{General verification protocol for optimal quantum channels}    \label{alg:generalprotocol}
    \Require{\Statex 
    \begin{itemize}
        \item $p_x$ \Comment{Probability distribution}
        \item classical description of~$\rho_x$ \Comment{Input states}
        \item classical description of~$\ket{\phi_x}\bra{\phi_x}$ \Comment{Output target states} 
        \item $\bar{F}_\text{t}$ \Comment{$0<\bar{F}_\text{t}<\bar{F}_{\mathrm{max}}$ is threshold average fidelity}
        \item $\delta$ \Comment{$0<\delta\le\frac{1}{2}$ is maximal failure probability}
        \item $\epsilon$ \Comment{$0<\epsilon <\frac{\bar{F}_{\mathrm{max}}-\bar{F}_\text{t}}{2}$ is error bound}
        \item $\mathcal E_{\mathrm{p}}$ \Comment{The sample complexity depends on both $\delta$ and $\epsilon$.}
         \item $\ket{\Psi}_{\mathrm{AR}}$ \Comment{The number of copies of~$\ket{\Psi}_{\mathrm{AR}}$ depends on that of~$\mathcal E_{\mathrm{p}}$.}
        \end{itemize}
    }
    \Ensure{\Statex
    \begin{itemize}
    \item $b$ \Comment{$b\in\{0, 1\}$, $0$ means reject and $1$ means accept.}
    \end{itemize}
    }

    \Procedure{VerificationofOptimalChannels}{$p_x$,~$x$, classical description of $\rho_x$ and $\ket{\phi_x}\bra{\phi_x}$, $\bar{F}_\text{t}$, $\delta$, $\epsilon$, $\mathcal E_{\mathrm{p}}$, $\ket{\Psi}_{\mathrm{AR}}$}
\State send system~A of each copy of~$\ket{\Psi}_{\mathrm{AR}}$ through one copy of~$\mathcal E_{\mathrm{p}}$;
\State apply local measurements on each~$\mathcal E_{\mathrm{p}}\otimes \mathcal I\left(\ket{\Psi}\bra{\Psi}_{\mathrm{AR}}\right)$ to measure~$W_{\mathrm{A}'\mathrm{R}}$;
\Statex
\Comment{$W_{\mathrm{A}'\mathrm{R}}$ is a tight lower bound of the observable $O_{\mathrm{A}'\mathrm{R}}$ in Eq.~(\ref{benchmarkObservable}).}
\State by processing measurement outcomes, obtain an estimate $\omega(\mathcal E_{\mathrm{p}})^*$ of $\omega(\mathcal E_{\mathrm{p}})$;
\Comment{With probability no less than~$1-\delta$, 
\begin{equation}
    \omega(\mathcal E_{\mathrm{p}})^*\in [\omega(\mathcal E_{\mathrm{p}})-\epsilon, \omega(\mathcal E_{\mathrm{p}})+\epsilon].
\end{equation}}
 \If{$\omega(\mathcal E_{\mathrm{p}})^* \ge \bar{F}_\text{t}+\epsilon$}
 \State \Return $b=1$;
 \Else \State \Return $b=0$.
 \EndIf
    \EndProcedure
    \end{algorithmic}
\end{algorithm}

This general verification protocol satisfy both the completeness and soundness conditions in definition~\ref{def:quantum-channel-verification}.
If $\mathcal E_{\mathrm{p}}$ is an optimal quantum channel, then $\omega(\mathcal E_{\mathrm{p}})=\bar{F}_{\mathrm{max}}$.
Hence, with probability at least $1-\delta$,
\begin{equation}
\omega(\mathcal E_{\mathrm{p}})^*\ge \bar{F}_{\mathrm{max}}-\epsilon 
> \bar{F}_\text{t}+2\epsilon-\epsilon
=\bar{F}_\text{t}+\epsilon.
\end{equation}
If $\bar{F}_{\mathcal E_{\mathrm{p}}}\le \bar{F}_\text{t}$, with probability at least $1-\delta$,
\begin{equation}
\omega(\mathcal E_{\mathrm{p}})^*\le \omega(\mathcal E_{\mathrm{p}})+\epsilon
< s_{\mathcal E_{\mathrm{p}}}+\epsilon
\le \bar{F}_\text{t}+\epsilon.
\end{equation}
Using the decision-making procedure, we conclude that this protocol satisfies the completeness and soundness conditions.

From the continuity of the function~$\omega(\mathcal E_p)$ at optimal quantum channels, 
a neighbourhood of optimal channels exists in the topological space of CPTP maps, such that $\forall\; \mathcal E_{\mathrm{p}}$ in this neighbourhood satisfies
\begin{equation}
    \omega(\mathcal E_{\mathrm{p}})\ge \bar{F}_\text{t}+2\epsilon.
\end{equation} 
Hence, with probability at least $1-\delta$,
\begin{equation}
    \omega(\mathcal E_{\mathrm{p}})^*\ge \bar{F}_\text{t}+\epsilon.
\end{equation} 
It indicates that
the verifier accepts any quantum channel in a neighbourhood of the optimal channels, with high probability, in the topological space.

This section has presented
our general scheme on how to verify
an optimal quantum channel in terms of average fidelity. We have mathematically defined optimal-quantum-channel verification and average-fidelity witness. In next section, we present examples of this general verification protocol by measuring experimentally feasible average-fidelity witnesses.

\section{Verification of bosonic channels}
\label{sec:result}

In this section, we present two verification protocols, one for multi-mode Gaussian unitary channels, the other for single-mode amplification channels.
All operations and sample complexities in the protocols are specified. The verification operations only require the preparation of two-mode squeezed vacuum states and the application of local homodyne detections.
The sample complexities scale polynomially with respect to all channel-specification parameters. In both protocols, we devise experimentally feasible average-fidelity witnesses, the mean values of which, can be sampled by local homodyne detections.

\subsection{\label{verifyGunitary}Verification of multi-mode Gaussian unitary channels}
In this subsection, we present a verification protocol for multi-mode Gaussian unitary channels. 
Central to this verification protocol, is an average-fidelity witness, and we show that the mean value of this witness
can be estimated by sampling the means and the covariance matrix of quadrature operators.

Here we investigate a verification protocol for the optimal quantum channel in terms of average fidelity
\begin{equation}
    \label{averageFidelityGaussianUnitary}
    \bar{F}(\mathcal E, \mathcal U_{\bm{S}, \bm{d}})\coloneqq\int \frac{\mathrm{d}^{2m} \bm{\alpha}}{\pi^m} \lambda^m \mathrm{e}^{-\lambda|\bm{\alpha}|^2} 
    \langle \bm{\alpha}| U_{\bm{S}, \bm{d}}^\dagger \mathcal E(\ket{\bm{\alpha}}\bra{\bm{\alpha}}) U_{\bm{S}, \bm{d}}\ket{\bm{\alpha}}\rangle,
\end{equation}
where
\begin{equation}
    \mathcal U_{\bm{S}, \bm{d}}(\rho)= U_{\bm{S}, \bm{d}}\rho U_{\bm{S}, \bm{d}}^\dagger,
\end{equation} 
is the unitary quantum channel and
\begin{equation}
\ket{\bm{\alpha}}\coloneqq\ket{\alpha_1}\otimes\ket{\alpha_2}\otimes\cdots\otimes\ket{\alpha_m},
\; \bm{\alpha}\coloneqq(\alpha_1,\alpha_2,\dots,\alpha_m)\in\mathbb{C}^{\otimes m}
\end{equation}
is a product of~$m$ coherent states. 
Evidently,
$\mathcal U_{\bm{S}, \bm{d}}$ achieves unity average fidelity~(\ref{averageFidelityGaussianUnitary}).

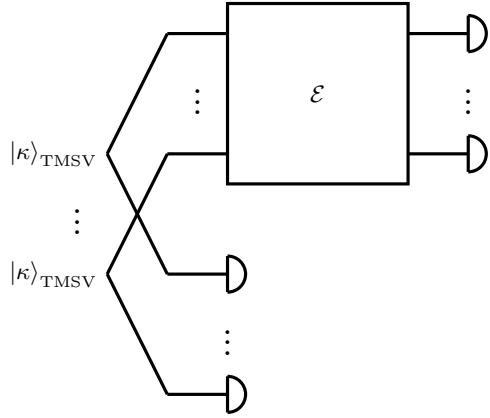
\begin{figure}[h]
    \centering
\begin{tikzpicture}[scale=0.8]
 
\draw[ very thick] (0,0) rectangle (3,3);
\draw[ very thick] (-1,0.5) -- (0,0.5); 
\draw[ very thick] (-1,2.5) -- (0,2.5); 
\draw[ very thick] (-1,-1.5) -- (0,-1.5); 
\draw[ very thick] (-1,-3.5) -- (0,-3.5); 
\draw[ very thick] (-1,2.5) -- (-2,0.5) node [anchor=east] {$\ket{\kappa}_{\mathrm{TMSV}}$}; 
\draw[ very thick] (-2,0.5) -- (-1,-1.5);  
\draw[ very thick] (-1,0.5) -- (-2,-1.5) node [anchor=east] {$\ket{\kappa}_{\mathrm{TMSV}}$};
\draw[ very thick] (-2,-1.5) -- (-1,-3.5);
\draw (-0.5, 1.5) node {\LARGE{$\vdots$}};
\draw (-2.5, -0.5) node {\LARGE{$\vdots$}};
\draw (0, -2.5) node {\LARGE{$\vdots$}};
\draw (1.5,1.5) node {$\mathcal E$};
\draw[ very thick] (3,0.5) -- (4,0.5); 
\draw[ very thick] (3,2.5) -- (4,2.5);
\draw[ very thick] (4,0.2) -- (4,0.8);
\draw[ very thick] (4,0.2) arc (-90:90:0.3);
\draw[ very thick] (4,2.2) -- (4,2.8);
\draw[ very thick] (4,2.2) arc (-90:90:0.3);
\draw (4, 1.5) node {\LARGE{$\vdots$}};
\draw[ very thick] (0,-1.8) -- (0,-1.2);
\draw[ very thick] (0,-1.8) arc (-90:90:0.3);
\draw[ very thick] (0,-3.8) -- (0,-3.2);
\draw[ very thick] (0,-3.8) arc (-90:90:0.3);
 
\end{tikzpicture}
    \caption{Our verification scheme for a multi-mode Gaussian unitary channel. Each $\ket{\kappa}_{\mathrm{TMSV}}$ denotes a two-mode squeezed vacuum state with squeezing parameter~$\kappa$. One mode of each~$\ket{\kappa}_{\mathrm{TMSV}}$ goes through a multi-mode unknown bosonic quantum channel, denoted by~$\mathcal E$ and represented by a square. Homodyne detections, represented by semicircles, are applied at each output mode of~$\mathcal E$ and the other mode of each~$\ket{\kappa}_{\mathrm{TMSV}}$.}
    \label{fig:multimodeGaussian}
    \end{figure}

The verification protocol for the optimal quantum channel in terms of the average fidelity~(\ref{averageFidelityGaussianUnitary}) is presented in Algorithm~\ref{alg:GaussianUnitaryOperation}. The schematic diagram of the verification scheme is shown in Fig.~\ref{fig:multimodeGaussian}. The protocol requires $2m c_3+m(2m+1) c_4+4m^2 c_5$ copies of~$\mathcal E_{\mathrm{p}}$,
where
\begin{align} \label{GaussianSampleComplexity1}
&c_3\in O\left(\frac{m^4\norm{\bm{S}}_{\infty}^4\norm{\bm{d}}^2\sigma_1^2}{ \varepsilon^2\ln(1/(1-\delta))}\right),\\  \label{GaussianSampleComplexity2}
            &c_4\in O\left( \frac{m^4 \norm{\bm{S}}_{\infty}^4\sigma_2^2}{\varepsilon^2 \ln(1/(1-\delta))}\right),\\
             \label{GaussianSampleComplexity3}
            &c_5\in O\left(\frac{m^4 \norm{\bm{S}}_{\infty}^2 \sigma_2^2}{\varepsilon^2 \ln(1/(1-\delta))}\right).
\end{align}
All the measurements in the protocol can be accomplished by $m+5$ local homodyne settings, and  the detailed measurement scheme is explained in Appendix~\ref{measurementscheme}.

\begin{algorithm}[H]
\caption{Verification protocol for multi-mode Gaussian unitary operations} \label{alg:GaussianUnitaryOperation}
    \begin{algorithmic}[1]    
    \Require{\Statex 
    \begin{itemize}
        \item $\frac{1}{\lambda}$ \Comment{Variance of the prior Gaussian distribution}
        \item $\bm{S}$  \Comment{$\bm{S}\in \mathrm{Sp}(2m,\mathbb{R})$}
        \item $\bm{d}$  \Comment{$\bm{d}\in \mathbb{R}^{2m}$}
        \item $\bar{F}_\text{t}$ \Comment{$0<\bar{F}_\text{t}<1$ is the threshold average fidelity}
        \item $\delta$ \Comment{$0<\delta\le\frac{1}{2}$ is the maximal failure probability}
        \item $\epsilon$ \Comment{$0<\epsilon< \frac{1-\bar{F}_\text{t}}{2}$ is the error bound}
        \item $\mathcal E_{\mathrm{p}}$ \Comment{$2m c_3+m(2m+1) c_4+4m^2 c_5$ copies of~$\mathcal E_{\mathrm{p}}$}
  \item $\ket{\kappa}_{\mathrm{TMSV}}$ \Comment{$2m^2 c_3+m^2(2m+1) c_4+ 4m^3 c_5$ copies of~$\ket{\kappa}_{\mathrm{TMSV}}$, where
  \begin{equation}\label{kappa}
        \kappa=\arctanh\frac{1}{\sqrt{\lambda+1}}
    \end{equation}}
 \item $\sigma_1$ \Comment{the upper bound of the variance of any $\hat{\bm{x}}^{\mathrm{A}'}_l$, $1\le l\le 2m$, on $\mathcal E_{\mathrm{p}}\otimes \mathcal I\left(\ket{\kappa}\bra{\kappa}_{\mathrm{TMSV}}^{\otimes m}\right)$.}
 \item $\sigma_2$ \Comment{the upper bound of the variance of any 
$\frac{1}{2}\left(\hat{\bm{x}}^{\mathrm{A}'}_u \hat{\bm{x}}^{\mathrm{A}'}_v+\hat{\bm{x}}^{\mathrm{A}'}_v\hat{\bm{x}}^{\mathrm{A}'}_u\right)$
and 
$\hat{\bm{x}}^{\mathrm{A}'}_u \hat{\bm{x}}^{\mathrm{R}}_v$
on $\mathcal E_{\mathrm{p}}\otimes \mathcal I\left(\ket{\kappa}\bra{\kappa}_{\mathrm{TMSV}}^{\otimes m}\right)$,
 where $1\le u, v\le 2m$.}
        \end{itemize}
    }

    \Ensure{\Statex
    \begin{itemize}
    \item $b$ \Comment{$b\in\{0, 1\}$, $0$ means reject and $1$ means accept.}
    \end{itemize}
    }

    \Procedure{VerificationofGaussianUnitaryOperations}{$\frac{1}{\lambda}$, $\bm{S}$, $\bm{d}$, $\bar{F}_\text{t}$, $\delta$, $\epsilon$, $\sigma_1$, $\sigma_2$, $\mathcal E_{\mathrm{p}}$, $\ket{\kappa}_{\mathrm{TMSV}}$}
\For{each copy of~$\mathcal E_{\mathrm{p}}$}
   \For{$j=1:m$}
      \State send one mode of one copy of~$\ket{\kappa}_{\mathrm{TMSV}}$ into $j$-input of~$\mathcal E_{\mathrm{p}}$;
      \State keep the other mode as a reference mode;
    \EndFor
\EndFor
 \For{$l=1:2m$}
 \For{$i=1:c_3$}\Comment{To estimate~
$  \bm{\gamma}\coloneqq \bar{\bm{x}}_{\mathrm{A}'} \in \mathbb{R}^{2m}$.}
\State apply a single-shot homodyne detection for quadrature $\hat{\bm{x}}^{\mathrm{A}'}_l$ on one copy of~$\mathcal E_{\mathrm{p}}\otimes \mathcal I\left(\ket{\kappa}\bra{\kappa}_{\mathrm{TMSV}}^{\otimes m}\right)$;
\EndFor
\State $\bm{\gamma}^*_l\leftarrow \frac{1}{c_3}\sum_{i=1}^{c_3} \chi^{\hat{\bm{x}}^{\mathrm{A}'}_l}_i$; 
 \Statex\Comment{$\bm{\gamma}^*$ is an estimate of $\bm{\gamma}$. $ \chi^{\hat{\bm{x}}^{\mathrm{A}'}_l}_i$ is $i$th measurement outcome with respect to quadrature $\hat{\bm{x}}^{\mathrm{A}'}_l$.}
 \For{$i=1:c_4$}\Comment{To estimate the diagonal elements in $\bm{\Gamma}_1 \coloneqq\left\langle  \hat{\bm{x}}_{\mathrm{A}'} \hat{\bm{x}}_{\mathrm{A}'}^\top\right\rangle \in \mathbb{R}^{2m\times 2m}$.}
\State apply a single-shot homodyne detection for quadrature $\hat{\bm{x}}^{\mathrm{A}'}_l $ on one copy of~$\mathcal E_{\mathrm{p}}\otimes \mathcal I\left(\ket{\kappa}\bra{\kappa}_{\mathrm{TMSV}}^{\otimes m}\right)$;

\EndFor
\State  $\left(\bm{\Gamma}_1^*\right)_{uu}\leftarrow \frac{1}{c_4}\sum_{i=1}^{c_4}\left(\chi^{\hat{\bm{x}}^{\mathrm{A}'}_u}_i\right)^2$;
\Comment{$\bm{\Gamma}_1^*$ is an estimate of
$   \bm{\Gamma}_{1}$.}

\EndFor

\For{$u=1:2m$} \Comment{To estimate the off-diagonal elements in $\bm{\Gamma}_1$.}
\For{$v=1:u-1$}
\If{$(u, v)\neq (2j, 2j-1)$ \textrm{ for } $j\in \{1, 2, \dots, m\}$}
\For{$i=1:c_4$}
\State apply two single-shot homodyne detections for quadratures 
$  \hat{\bm{x}}^{\mathrm{A}'}_u $ and
$  \hat{\bm{x}}^{\mathrm{A}'}_v $ simultaneously
 on one copy of~$\mathcal E_{\mathrm{p}}\otimes \mathcal I\left(\ket{\kappa}\bra{\kappa}_{\mathrm{TMSV}}^{\otimes m}\right)$;
 \EndFor
 \State $\left(\bm{\Gamma}_1^*\right)_{uv}\leftarrow \frac{1}{c_4}\sum_{i=1}^{c_4}\chi^{\hat{\bm{x}}^{\mathrm{A}'}_u}_i \chi^{\hat{\bm{x}}^{\mathrm{A}'}_v}_i$;
\Else
\For{$i=1:c_4$}
\State apply a single-shot homodyne detection for quadrature 
$ \frac{1}{\sqrt{2}}\left( \hat{\bm{x}}^{\mathrm{A}'}_u+\hat{\bm{x}}^{\mathrm{A}'}_v\right)  $ on one copy of~$\mathcal E_{\mathrm{p}}\otimes \mathcal I\left(\ket{\kappa}\bra{\kappa}_{\mathrm{TMSV}}^{\otimes m}\right)$;
\EndFor
\State $\left(\bm{\Gamma}_1^*\right)_{uv}\leftarrow \frac{1}{c_4}\sum_{i=1}^{c_4}\left(\chi^{\frac{1}{\sqrt{2}}\left( \hat{\bm{x}}^{\mathrm{A}'}_u+\hat{\bm{x}}^{\mathrm{A}'}_v\right)}_i \right)^2- \frac{1}{2}(\bm{\gamma}^*_u)^2- \frac{1}{2}(\bm{\gamma}^*_v)^2$; 
\EndIf
  \State $\left(\bm{\Gamma}_1^*\right)_{vu}\leftarrow  \left(\bm{\Gamma}_1^*\right)_{uv}$;
\EndFor
\EndFor

    \algstore{GaussianUnitaryAlgPart1}
    \end{algorithmic}
\end{algorithm}
\begin{algorithm}[H]
 \begin{algorithmic}[1]
\algrestore{GaussianUnitaryAlgPart1}

\For{$u=1:2m$} \Comment{To estimate $    \bm{\Gamma}_{2}\coloneqq \left\langle  \hat{\bm{x}}_{\mathrm{A}'} \hat{\bm{x}}_{\mathrm{R}}^\top\right\rangle \in \mathbb{R}^{2m\times 2m}$.}
\For{$v=1:2m$}
\For{$i=1:c_5$}
\State apply two single-shot homodyne detection for
$\hat{\bm{x}}^{\mathrm{A}'}_u $ and $\hat{\bm{x}}^{\mathrm{R}}_v$ simultaneously on one copy of~$\mathcal E_{\mathrm{p}}\otimes \mathcal I\left(\ket{\kappa}\bra{\kappa}_{\mathrm{TMSV}}^{\otimes m}\right)$; 
\EndFor
\State $\left(\bm{\Gamma}_2^*\right)_{uv}\leftarrow \frac{1}{c_5}\sum_{i=1}^{c_5} \chi^{\hat{\bm{x}}^{\mathrm{A}'}_u }_i
\chi^{\hat{\bm{x}}^{\mathrm{R}}_v}_i$;
\Comment{$\bm{\Gamma}_2^*$ is an estimate of $\bm{\Gamma}_{2}$.}
 \State $\left(\bm{\Gamma}_2^*\right)_{vu}\leftarrow  \left(\bm{\Gamma}_2^*\right)_{uv}$;
\EndFor
\EndFor

\State 
$  \omega_{U_{\bm{S}, \bm{d}}}(\mathcal E_{\mathrm{p}})^*\leftarrow
    -\frac{1}{2}\tr \left[\bm{S}^{-\mathrm{T}}\bm{S}^{-1} \left(\bm{\Gamma}_1^* -2\bm{\gamma}^*\bm{d}^\top+\bm{d}\bm{d}^\top\right) \right]
+\frac{1}{\sqrt{\lambda+1}}\tr\left(\bm{Z}^{\oplus m} \bm{S}^{-1} \bm{\Gamma}_2^*\right) + \frac{m(\lambda^2-2\lambda-4)}{2\lambda(\lambda+1)}+1$;
\Statex 
\Comment{Obtain an estimate~$\omega_{U_{\bm{S}, \bm{d}}}(\mathcal E_{\mathrm{p}})^*$ of
        $\omega_{U_{\bm{S}, \bm{d}}}(\mathcal E_{\mathrm{p}})$
        in Eq.~(\ref{witnessUnitaryEstimator}).}
 \If{$\omega_{U_{\bm{S}, \bm{d}}}(\mathcal E_{\mathrm{p}})^* \ge \bar{F}_\text{t}+\epsilon$}
 \State \Return $b=1$;
 \Else \State \Return $b=0$.
 \EndIf
    \EndProcedure
    \end{algorithmic}
\end{algorithm}

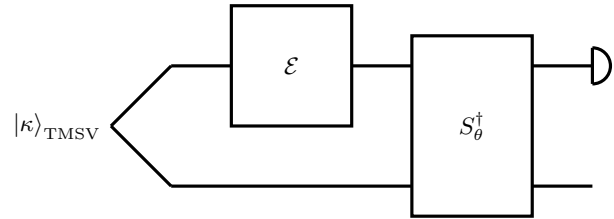
\begin{figure}[h]
\centering
    \begin{tikzpicture}[scale=0.8]
    \draw[very thick] (0,0) rectangle (2,2);
    \draw[ very thick] (-1,1) -- (0,1); 
    \draw[ very thick] (-1,-1) -- (3,-1); 
    \draw[ very thick] (-1,1) -- (-2,0) node [anchor=east] {$\ket{\kappa}_{\mathrm{TMSV}}$}; 
    \draw[ very thick] (-2,0) -- (-1,-1);  
    \draw (1,1) node {$\mathcal E$};
    \draw[ very thick] (3,-1.5) rectangle (5,1.5);
    \draw[ very thick] (2,1) -- (3,1);
    \draw (4,0) node {$S_\theta^\dagger$};
    \draw[ very thick] (5,1) -- (6,1);
    \draw[ very thick] (5,-1) -- (6,-1);
    \draw[ very thick] (6,1.3) -- (6,0.7);
    \draw[ very thick] (6,1.3) arc (90:-90:0.3);
    \end{tikzpicture}
    \caption{Previous benchmarking scheme for a single-mode bosonic amplification/attenuation channel~\cite{bai2018test}. $\ket{\kappa}_{\mathrm{TMSV}}$ denotes a two-mode squeezed vacuum state with squeezing parameter~$\kappa$. One mode of~$\ket{\kappa}_{\mathrm{TMSV}}$ goes through~$\mathcal E$. 
    The square, denoted by~$\mathcal E$, represents a single-mode unknown bosonic quantum channel. The output mode of~$\mathcal E$ and the other mode of~$\ket{\kappa}_{\mathrm{TMSV}}$ go through an online two-squeezing operation, denoted by~$S_\theta^\dagger$ and represented by a rectangle. A heterodyne detection, represented by a semicircle, is applied at one final output mode, and the other output mode is discarded.}
    \label{fig:singlemode}
\end{figure}

Now we devise an average-fidelity witness for the average fidelity in Eq.~(\ref{averageFidelityGaussianUnitary})
and show that its mean value is a linear combination of~$\bm{\gamma}$,~$\bm{\Gamma}_1$ and~$\bm{\Gamma}_2$.
Hence, the mean value of the witness
can be estimated by the measurement and classical-information processing schemes in Algorithm~\ref{alg:GaussianUnitaryOperation}.
\begin{theorem}\label{theorem:GaussianUnitary}
The observable
 \begin{equation}\label{averageFidelityWitnessGaussianUnitary}
\mathds{1}-\frac{\lambda}{\lambda+1} U_{\bm{S},\bm{d}}\otimes \mathds{1} \left(\sum_{i=1}^m  S_{\kappa} \hat{n}_i \otimes \mathds{1}  S_{\kappa}^\dagger \right)U_{\bm{S},\bm{d}}^\dagger\otimes \mathds{1}
\end{equation} 
is an average-fidelity witness for~$\bar{F}(\mathcal E, \mathcal U_{\bm{S}, \bm{d}})$ on~$\mathcal E\otimes \mathcal I \left(\ket{\kappa}\bra{\kappa}_{\mathrm{TMSV}}^{\otimes m}\right)$.
\end{theorem}
\noindent From now on, we use $W_{U_{\bm{S}, \bm{d}}}$ to denote the average-fidelity witness~(\ref{averageFidelityWitnessGaussianUnitary}). To show Theorem~\ref{theorem:GaussianUnitary}, we need Lemmas~\ref{lemma:AverageFidelityObservable} and~\ref{lemma:GaussianFidelityWitness}.

\begin{lemma}\label{lemma:AverageFidelityObservable}
Given performance operator
\begin{equation}\label{performanceOperatorAmplification}
\Omega_{\mathrm{A'A}}=\int \frac{d^2\alpha}{\pi}\lambda \mathrm{e}^{-\lambda|\alpha|^2} \ket{g\alpha}\bra{g\alpha}\otimes\ket{\alpha}\bra{\alpha},
\end{equation}
where $g>0$,
and input state~$\ket{\Psi}_{\mathrm{AR}}=\ket{\kappa}_{\mathrm{TMSV}}$,
if $g\le \sqrt{\lambda+1}$,
then
\begin{equation}\label{observableIdentityAttenuation}
    O_{\mathrm{A}'\mathrm{R}}=S_{\theta}(G_{\theta}\otimes \mathds{1})S_{\theta}^\dagger ,
\end{equation}
where 
\begin{equation}\label{GthetaOperator}
G_\theta =\sum_{n=0}^\infty \tanh^{2n}\theta\ket{n}\bra{n}
\end{equation}
and
\begin{equation}\label{TanhTheta}
    \theta=\arctanh\frac{g}{\sqrt{\lambda+1}};
\end{equation}
otherwise, 
\begin{equation}\label{observableAmplification}
   O_{\mathrm{A}'\mathrm{R}}=\tanh^2\theta' S_{\theta'}(\mathds{1}\otimes G_{\theta'})S_{\theta'}^\dagger,
\end{equation}
where 
\begin{equation}\label{TanhThetaPrime}
    \theta'=\arctanh\frac{\sqrt{\lambda+1}}{g}.
\end{equation}

\end{lemma}
\noindent
Ref.~\cite{bai2018test} has shown the results in Lemma~\ref{lemma:AverageFidelityObservable}, except missing the constant~$\tanh^2\theta'$ in Eq.~(\ref{observableAmplification}). 
The proof of this lemma is in Appendix~\ref{app:observable}. 

Lemma~\ref{lemma:AverageFidelityObservable} implies that by applying two-mode squeezing and measuring~$G_{\theta}$ at one mode, the verifier can directly estimate the average fidelity. As
\begin{equation}
    G_\theta=\coth^2\theta \int \frac{\text{d}^2 \alpha }{\pi} \mathrm{e}^{-\frac{|\alpha|^2}{\sinh^2\theta}}\ket{\alpha}\bra{\alpha},
\end{equation}
the mean value of $G_\theta$ can be estimated by using heterodyne detections~\cite{bai2018test}. This benchmark scheme also requires quantum memory to keep the entanglement between the output mode and the reference mode, and online two-mode squeezing to squeeze the combination of an unknown quantum state at the output mode and a thermal state at the reference mode. The schematic diagram of this method, devised in~\cite{bai2018test}, is shown in Fig.~\ref{fig:singlemode}. However, the combination of quantum memory, online squeezing and heterodyne detections is experimentally challenging. 

To devise an experimentally feasible verification scheme, we find lower bounds of the observables in Lemma~\ref{lemma:AverageFidelityObservable} using the lemma below.
\begin{lemma}\label{lemma:GaussianFidelityWitness}
For
any $\theta>0$, $m\in \mathbb{N}^+$,
\begin{equation}\label{lemmaInequality}
    G_\theta^{\otimes m}
\ge \mathds{1}-\frac{\sum_{i=1}^m\hat{n}_i}{\cosh^2\theta}.
\end{equation}
\end{lemma}
\noindent
As far as we know, the inequality in Lemma~\ref{lemma:GaussianFidelityWitness} is novel and has not appeared in any previous literatures.
The proof of this lemma is in Appendix~\ref{proofLemma}. 
Combining Lemma~\ref{lemma:GaussianFidelityWitness} with Lemma~\ref{lemma:AverageFidelityObservable}, we obtain the observable in Eq.~(\ref{averageFidelityWitnessGaussianUnitary}). 
Now we prove Theorem~\ref{theorem:GaussianUnitary}.
\begin{proof}
From Eq.~(\ref{average-fidelity-based-performance-operator}), we know that
the performance operator,
in the test of average fidelity~$\bar{F}(\mathcal E, \mathcal U_{\bm{S}, \bm{d}})$, is
\begin{equation}
\Omega_{\mathrm{A'A}}
    =\int \frac{\mathrm{d}^{2m} \bm{\alpha}}{\pi^m} \lambda^m \mathrm{e}^{-\lambda|\bm{\alpha}|^2}
U_{\bm{S},\bm{d}}\ket{\bm{\alpha}}\bra{\bm{\alpha}}U_{\bm{S},\bm{d}}^\dagger\otimes \ket{\bm{\alpha}}\bra{\bm{\alpha}}.
\end{equation}
Using Eq.~(\ref{observableIdentityAttenuation}) for the tensor product of~$m$ modes, we obtain the observable
\begin{equation}\label{averageScoreGaussianUnitary}
    O_{\mathrm{A}'\mathrm{R}}= U_{\bm{S},\bm{d}}\otimes \mathds{1}  S_{\kappa}^{\otimes m} G_\kappa^{\otimes m}\otimes \mathds{1}  S_{\kappa}^{\dagger \otimes m} U_{\bm{S},\bm{d}}^\dagger\otimes \mathds{1},
\end{equation}
such that
\begin{equation}\label{FideilityObservableGaussianUnitary}
\bar{F}(\mathcal E, \mathcal U_{\bm{S}, \bm{d}})=\tr \left[O_{\mathrm{A}'\mathrm{R}}\mathcal E\otimes \mathcal I\left(\ket{\kappa}\bra{\kappa}_{\mathrm{TMSV}}^{\otimes m}\right) \right].
\end{equation}
In Eq.~(\ref{averageScoreGaussianUnitary}),  each $G_{\kappa}$ acts on one output mode, each $S_{\kappa}$ acts on one output mode and the associated reference mode, and $U_{\bm{S}, \bm{d}}$ acts on the~$m$ output modes.
To perform the operator multiplication in Eq.~(\ref{averageScoreGaussianUnitary}),
the operators must be represented on the Hilbert spaces with one specific order,
like $\mathrm{A}'_1, \dots, \mathrm{A}'_m, \mathrm{R}_1, \dots, \mathrm{R}_m$.
The permutation of Hilbert spaces leave the operators unchanged.

Plugging inequality~(\ref{lemmaInequality}) into Eqs.~(\ref{averageScoreGaussianUnitary}) and~(\ref{FideilityObservableGaussianUnitary}) yields
\begin{equation}
\bar{F}(\mathcal E, \mathcal U_{\bm{S}, \bm{d}})\ge\tr\left[W_{U_{\bm{S}, \bm{d}}} \mathcal E\otimes \mathcal I\left(\ket{\kappa}\bra{\kappa}_{\mathrm{TMSV}}^{\otimes m}\right)\right],
\end{equation}
which proves condition~(\ref{averagewitnessRequirement2}).
On the other hand, from Eqs.~(\ref{averageScoreGaussianUnitary}) and~(\ref{FideilityObservableGaussianUnitary}),
we have
\begin{equation}
    \bar{F}(\mathcal E, \mathcal U_{\bm{S}, \bm{d}})=\tr\left\{G_\theta \tr_{\mathrm{R}}\left[ S_{\kappa}^{\dagger \otimes m} U_{\bm{S},\bm{d}}^\dagger\otimes \mathds{1}\mathcal E\otimes \mathcal I\left(\ket{\kappa}\bra{\kappa}_{\mathrm{TMSV}}^{\otimes m}\right)U_{\bm{S},\bm{d}}\otimes \mathds{1}  S_{\kappa}^{\otimes m} \right] \right\}.
\end{equation}
Using Eq.~(\ref{GthetaOperator}), we know that~$\mathcal E$ is an optimal channel, i.e., $\bar{F}(\mathcal E, \mathcal U_{\bm{S}, \bm{d}})$ achieves one, iff
\begin{equation}
    \tr_{\mathrm{R}}\left[ S_{\kappa}^{\dagger \otimes m} U_{\bm{S},\bm{d}}^\dagger\otimes \mathds{1}\mathcal E\otimes \mathcal I\left(\ket{\kappa}\bra{\kappa}_{\mathrm{TMSV}}^{\otimes m}\right)U_{\bm{S},\bm{d}}\otimes \mathds{1}  S_{\kappa}^{\otimes m} \right]=\ket{0}\bra{0}^{\otimes m},
\end{equation}
which is further equivalent to
\begin{equation}
\tr \left[ W_{U_{\bm{S}, \bm{d}}}\mathcal E\otimes \mathcal I\left(\ket{\kappa}\bra{\kappa}_{\mathrm{TMSV}}^{\otimes m}\right)\right]=1.
\end{equation}
This proves condition~(\ref{averagewitnessRequirement1}).
Thus, $W_{U_{\bm{S}, \bm{d}}}$ is an average-fidelity witness for~$\bar{F}(\mathcal E, \mathcal U_{\bm{S}, \bm{d}})$.
\end{proof}

Next we show that  the expectation value of the average-fidelity witness
\begin{equation}
\omega_{U_{\bm{S}, \bm{d}}}(\mathcal E_{\mathrm{p}}) \coloneqq
        \tr \left[ W_{U_{\bm{S}, \bm{d}}}\mathcal E_{\mathrm{p}}\otimes \mathcal I\left(\ket{\kappa}\bra{\kappa}_{\mathrm{TMSV}}^{\otimes m}\right)\right]
\end{equation}
 is a linear combination of the mean values of quadrature operators,~$\bm{\gamma}$, and the covariances of quadrature operators,~$\bm{\Gamma}_1$ and~$\bm{\Gamma}_2$.
We rewrite each photon number operator in terms of position and momentum operators,
\begin{equation}
 \label{photonNumberToQuadrature}
\hat{n}=\frac{\hat{\bm{x}}^\top\hat{\bm{x}}-m}{2}.
\end{equation}
By applying the inverse transformations of~(\ref{two-modeSqueezingTransformation})  
\begin{equation}\label{HeisenbergPictureSqueezingOperation}
    S_{\kappa}^{\otimes m} 
    \begin{bmatrix}
    \hat{\bm{x}}_{\mathrm{A}'} \\
    \hat{\bm{x}}_{\mathrm{R}}
    \end{bmatrix}
    S_{\kappa}^{\otimes m\dagger}=
    \begin{bmatrix}
    \cosh\kappa \bm{\mathds{1}}^{\oplus m} &
    -\sinh\kappa \bm{Z}^{\oplus m}\\
    -\sinh\kappa \bm{Z}^{\oplus m} & 
    \cosh\kappa \bm{\mathds{1}}^{\oplus m}
    \end{bmatrix}
    \begin{bmatrix}
    \hat{\bm{x}}_{\mathrm{A}'} \\
    \hat{\bm{x}}_{\mathrm{R}}
    \end{bmatrix},
\end{equation}
and the inverse transformation of~(\ref{GaussianUnitary})
\begin{equation}
 U_{\bm{S},\bm{d}}  \hat{\bm{x}}_{\mathrm{A}'} U_{\bm{S},\bm{d}}^\dagger
=   \bm{S}^{-1} (  \hat{\bm{x}}_{\mathrm{A}'}- \bm{d}),
\end{equation}
we write~$W_{U_{\bm{S}, \bm{d}}}$ in terms of~$\hat{\bm{x}}_{\mathrm{A}'}$ and~$\hat{\bm{x}}_{\mathrm{R}}$,
\begin{equation}
    W_{U_{\bm{S}, \bm{d}}}=\cosh^2\kappa (\hat{\bm{x}}_{\mathrm{A}'}^\top-\bm{d}^\top)\bm{S}^{-\mathrm{T}}\bm{S}^{-1}(\hat{\bm{x}}_{\mathrm{A}'}-\bm{d})-\sinh(2\kappa)\hat{\bm{x}}_{\mathrm{R}}^\top
    \bm{Z}^{\oplus m}\bm{S}^{-1}(\hat{\bm{x}}_{\mathrm{A}'}-\bm{d})+\sinh^2\kappa\hat{\bm{x}}_{\mathrm{R}}^\top\hat{\bm{x}}_{\mathrm{R}}.
\end{equation}
As each reference mode is in a thermal state $\rho_T(\frac{1}{\lambda})$, for each $1\le l\le 2m$,
\begin{equation}\label{referenceModeThermal}
\left\langle\left(\hat{\bm{x}}_l^\mathrm{R}\right)^2\right\rangle=\frac{\lambda+2}{\lambda}.
\end{equation}
Using this fact and Eq.~(\ref{kappa}), we obtain 
\begin{equation}
\label{witnessUnitaryEstimator}
\omega_{U_{\bm{S}, \bm{d}}}(\mathcal E_{\mathrm{p}}) =-\frac{1}{2}\tr \left[\bm{S}^{-\mathrm{T}}\bm{S}^{-1} \left(\bm{\Gamma}_1 -2\bm{\gamma}\bm{d}^\top+\bm{d}\bm{d}^\top\right) \right] 
+\frac{1}{\sqrt{\lambda+1}}\tr\left(\bm{Z}^{\oplus m} \bm{S}^{-1} \bm{\Gamma}_2\right) + \frac{m(\lambda^2-2\lambda-4)}{2\lambda(\lambda+1)}+1.
\end{equation}
Eq.~(\ref{witnessUnitaryEstimator}) implies that the mean value of the average-fidelity witness can be estimated by sampling the means and the covariance matrix of quadrature operators, as shown in Algorithm~\ref{alg:GaussianUnitaryOperation}.

This subsection has presented a verification protocol for multi-mode Gaussian unitary channels including all operations and sample complexities.
Central to the verification protocol,
we have devised an average-fidelity witness and show that its mean value can be estimated by applying local homodyne detections.
Our protocol greatly simplifies the experimental setting to detect the average fidelity without requiring quantum memory or online squeezing. 
The sample complexity of this protocol scales polynomially with the number of modes, the maximal squeezing parameter and the phase-space displacement of the target Gaussian unitary operation. 

\subsection{Verification of single-mode amplification channels}
\label{verifyamplifyattenuation}
In this subsection, we present a verification protocol for single-mode amplification channels. We devise an average-fidelity witness for this verification protocol and show that its mean value is a linear combination of the covariances of quadrature operators.

Quantum amplification channels~\cite{pooser2009low} are important for quantum cloning and other quantum information processing protocols. We investigate a verification protocol for the optimal quantum channel in terms of average fidelity
\begin{equation}
    \label{averageFidelityAmplification}
    \bar{F}_g(\mathcal E)=\int \frac{\mathrm{d}^{2} \alpha}{\pi} \lambda \mathrm{e}^{-\lambda|\alpha|^2} 
    \bra{g\alpha} \mathcal E(\ket{\alpha}\bra{\alpha}) \ket{g \alpha},
\end{equation}
where $g>\lambda+1$ is the amplification gain. 
Chiribella and Xie showed that the optimal amplification channel can be achieved by a Gaussian amplification channel,
using two-mode squeezing, and the maximum achievable average fidelity~(\ref{averageFidelityAmplification}) is~\cite{chiribella2013optimal}
\begin{equation}
    \bar{F}_g^{\mathrm{max}}=\frac{\lambda+1}{g^2}.
\end{equation}
We present our verification protocol in Algorithm~\ref{alg:amplificationchannel}. The protocol requires $2c_6+2c_7$  copies of~$\mathcal E_{\mathrm{p}}$,
where
\begin{equation}\label{amplificationSampleComplexity1}
           c_6\in O\left(\frac{ g^4 \sigma_2^2}{\varepsilon^2\ln(1/(1-\delta))}\right)
       \end{equation}
       and \begin{equation}\label{amplificationSampleComplexity2}
           c_7\in O\left(\frac{g^6\sigma_2^2}{\varepsilon^2\ln(1/(1-\delta))}\right).
       \end{equation}
Thus, the sample complexity scales efficiently with respect to amplification gain~$g$.

\begin{algorithm}[H]
 \caption{Verification protocol for single-mode amplification channel}\label{alg:amplificationchannel}
    \begin{algorithmic}[1]    
    \Require{\Statex 
    \begin{itemize}
        \item $\frac{1}{\lambda}$ \Comment{Variance of the prior Gaussian distribution}
        \item $g$  \Comment{$g>\lambda+1$ is the amplification gain.}
        \item $\bar{F}_\text{t}$ \Comment{$0<\bar{F}_\text{t}<\frac{\lambda+1}{g^2}$ is the threshold average fidelity.}
        \item $\delta$ \Comment{$0<\delta\le\frac{1}{2}$ is the maximal failure probability.}
        \item $\epsilon$ \Comment{$0<\epsilon< \frac{\lambda+1-g^2\bar{F}_\text{t}}{2g^2}$ is the error bound.}
        \item $\mathcal E_{\mathrm{p}}$ \Comment{$2c_6+2c_7$ copies of~$\mathcal E_{\mathrm{p}}$ from the prover}
  \item $\ket{\kappa}_{\mathrm{TMSV}}$ \Comment{$2c_6+2c_7$ copies of~$\ket{\kappa}_{\mathrm{TMSV}}$}
 \item $\sigma_2$ \Comment{the upper bound of the variances of $\hat{q}^2_{\mathrm{A}'}$, $\hat{p}^2_{\mathrm{A}'}$, $\hat{q}_{\mathrm{A}'}\hat{q}_{\mathrm{R}}$ and $\hat{p}_{\mathrm{A}'}\hat{p}_{\mathrm{R}}$ on $\mathcal E_{\mathrm{p}}\otimes \mathcal I\left(\ket{\kappa}\bra{\kappa}_{\mathrm{TMSV}}\right)$.}
        \end{itemize}
    }
    \Ensure{\Statex
    \begin{itemize}
    \item $b$ \Comment{$b\in\{0, 1\}$, $0$ means reject and $1$ means accept.}
    \end{itemize}
    }

    \Procedure{VerificationofAmplificationChannel}{$\frac{1}{\lambda}$, $g$, $\bar{F}_\text{t}$, $\delta$, $\epsilon$, $\sigma_2$, $\mathcal E_{\mathrm{p}}$, $\ket{\kappa}_{\mathrm{TMSV}}$}
\State send one mode of each copy of $\ket{\kappa}_{\mathrm{TMSV}}$ into a copy of~$\mathcal E_{\mathrm{p}}$, and keep the other mode as a reference mode;
\For{$i=1:c_6$}
\State apply a single-shot homodyne detection for quadrature $\hat{q}_{\mathrm{A}'}$ on one copy
       of~$\mathcal E_{\mathrm{p}}\otimes\mathcal I\left(\ket{\kappa}\bra{\kappa}_{\mathrm{TMSV}}\right)$;
 \EndFor
\State $\braket{\hat{q}_{\mathrm{A}'}^2}^*\leftarrow\frac{1}{c_6}\sum_{i=1}^{c_6}\left(\chi^{\hat{q}_{\mathrm{A}'}}_i\right)^2$;
 \Comment{$\braket{\hat{q}_{\mathrm{A}'}^2}^*$ is an estimate of $\braket{ \hat{q}_{\mathrm{A}'}^2}$.}
 \For{$i=1:c_6$}
 \State apply a single-shot homodyne detection for quadrature $\hat{p}_{\mathrm{A}'}$ on one copy 
        of~$\mathcal E_{\mathrm{p}}\otimes\mathcal I\left(\ket{\kappa}\bra{\kappa}_{\mathrm{TMSV}}\right)$;
 \EndFor
\State $\braket{\hat{p}_{\mathrm{A}'}^2}^*\leftarrow \frac{1}{c_6}\sum_{i=1}^{c_6}\left(\chi^{\hat{p}_{\mathrm{A}'}}_i\right)^2$;
      \Comment{$\braket{\hat{p}_{\mathrm{A}'}^2}^*$ is an estimate of $\braket{\hat{p}_{\mathrm{A}'}^2}$.}
\For{$i=1:c_7$}
\State apply two single-shot homodyne detections for quadratures
       $\hat{q}_{\mathrm{A}'}$ and $\hat{q}_{\mathrm{R}}$ simultaneously on one copy of~$\mathcal E_{\mathrm{p}}\otimes\mathcal I\left(\ket{\kappa}\bra{\kappa}_{\mathrm{TMSV}}\right)$; 
\EndFor
 \State $\braket{ \hat{q}_{\mathrm{A}'}\hat{q}_{\mathrm{R}}}^*\leftarrow \frac{1}{c_7}\sum_{i=1}^{c_7}\chi_i^{\hat{q}_{\mathrm{A}'}}\chi_i^{\hat{q}_{\mathrm{R}}}$;
 \Comment{$\braket{ \hat{q}_{\mathrm{A}'}\hat{q}_{\mathrm{R}}}^*$ is an estimate of $\braket{ \hat{q}_{\mathrm{A}'}\hat{q}_{\mathrm{R}}}$.}
 \For{$i=1:c_7$}
 \State apply two single-shot homodyne detections for quadratures $\hat{p}_{\mathrm{A}'}$ and $\hat{p}_{\mathrm{R}}$ simultaneously on one copy of~$\mathcal E_{\mathrm{p}}\otimes\mathcal I\left(\ket{\kappa}\bra{\kappa}_{\mathrm{TMSV}}\right)$; 
 \EndFor
 \State $\braket{ \hat{p}_{\mathrm{A}'}\hat{p}_{\mathrm{R}}}^*\leftarrow \frac{1}{c_7}\sum_{i=1}^{c_7}\chi_i^{\hat{p}_{\mathrm{A}'}}\chi_i^{\hat{p}_{\mathrm{R}}}$; 
 \Comment{$\braket{ \hat{p}_{\mathrm{A}'}\hat{p}_{\mathrm{R}}}^*$ is an estimate of $\braket{ \hat{p}_{\mathrm{A}'}\hat{p}_{\mathrm{R}}}$.}
\State $\omega(\mathcal E_{\mathrm{p}})^*\leftarrow
\frac{g^2}{\lambda+1}\left[\frac{3}{2}-\frac{1}{2} \left(\braket{\hat{q}_{\mathrm{A}'}^2}^*+\braket{\hat{p}_{\mathrm{A}'}^2}^*\right)+\frac{g}{\sqrt{\lambda+1}} \left(\braket{ \hat{q}_{\mathrm{A}'}\hat{q}_{\mathrm{R}}}^*-\braket{ \hat{p}_{\mathrm{A}'}\hat{p}_{\mathrm{R}}}^*\right)- \frac{g^2(3\lambda+4)}{2\lambda(\lambda+1)}\right]$
\Statex 
\Comment{Obtain an estimate $\omega(\mathcal E_{\mathrm{p}})^*$ of~$\omega(\mathcal E_{\mathrm{p}})$
        in Eq.~(\ref{AmplifcationWitnessInQuadratureBasis}).}
 \If{$\omega(\mathcal E_{\mathrm{p}})^* \ge \bar{F}_\text{t}+\epsilon$}
 \State \Return $b=1$;
 \Else \State \Return $b=0$.
 \EndIf
    \EndProcedure
    \end{algorithmic}
\end{algorithm}

Central to our verification protocol, we devise an average-fidelity witness and show that its mean value can be estimated by the measurement and classical-information processing scheme in Algorithm~\ref{alg:amplificationchannel}. 

\begin{theorem} \label{thm:amplificationchannel}
The observable
  \begin{equation}\label{averageFidelityWitnessAmplification}
     \frac{\lambda+1}{g^2} \left(\mathds{1}-\frac{g^2-\lambda-1}{g^2}S_{\theta'} \mathds{1}\otimes \hat{n} S_{\theta'}^\dagger\right)
\end{equation}
is an average-fidelity witness for $\bar{F}_g(\mathcal E)$ on~$\mathcal E\otimes \mathcal I\left(\ket{\kappa}\bra{\kappa}_{\mathrm{TMSV}}\right)$.
\end{theorem}
\noindent
Henceforth, we use $W_{\mathrm{amp}}$ to denote the average-fidelity witness~(\ref{averageFidelityWitnessAmplification}).
Lemma~\ref{lemma:AverageFidelityObservable} implies that the average fidelity of an amplification channel can be estimated 
by applying quantum memory, online two-mode squeezing and heterodyne detections as shown in Fig.~\ref{fig:singlemode}.  
However, this method is experimentally challenging. Measuring the average-fidelity witness in Theorem~\ref{thm:amplificationchannel} provides an experimentally feasible method.

\begin{proof}
From Eq.~(\ref{observableAmplification}), we know
\begin{equation}\label{averageFidelityAmplificationSqueezedInput}
    \bar{F}_g(\mathcal E)=\frac{\lambda+1}{g^2} \tr \left[S_{\theta'} \mathds{1}\otimes G_{\theta'} S_{\theta'}^\dagger \mathcal E\otimes \mathcal I \left(\ket{\kappa}\bra{\kappa}_{\mathrm{TMSV}}\right) \right].
\end{equation}
Plugging in inequality~(\ref{lemmaInequality}), we have
\begin{equation}\label{amplificationChannelWitnessCondition2}
    \forall \mathcal E, \; \tr \left[W_{\mathrm{amp}} \mathcal E\otimes \mathcal I \left(\ket{\kappa}\bra{\kappa}_{\mathrm{TMSV}}\right)\right] \le \bar{F}_g(\mathcal E),
\end{equation}
which proves condition~(\ref{averagewitnessRequirement2}).
On the other hand, from Eqs.~(\ref{GthetaOperator}) and (\ref{averageFidelityAmplificationSqueezedInput}), we know that $\mathcal E$ is optimal; i.e.,  $\bar{F}_g(\mathcal E)=\frac{\lambda+1}{g^2}$, iff 
\begin{equation}\label{traceoutVacuumAmplification}
    \tr_{\mathrm{A}'} \left[S_{\theta'}^\dagger\mathcal E\otimes \mathcal I \left(\ket{\kappa}\bra{\kappa}_{\mathrm{TMSV}}\right) S_{\theta'}\right]=\ket{0}\bra{0}.
\end{equation}
Eq.~(\ref{traceoutVacuumAmplification}) is further equivalent to
\begin{equation}
   \tr\left[W_{\mathrm{amp}} \mathcal E\otimes \mathcal I (\ket{r}\bra{r})\right]= \frac{\lambda+1}{g^2},
\end{equation}
which proves condition~(\ref{averagewitnessRequirement1}).
Thus,  we conclude that
$W_{\mathrm{amp}}$ is an average-fidelity witness for $\bar{F}_g(\mathcal E)$.
\end{proof}

Next we show that the expectation value of the average-fidelity witness
\begin{equation}
    \omega_{\mathrm{amp}}(\mathcal E_{\mathrm{p}})\coloneqq\tr\left[W_{\mathrm{amp}} \mathcal E_{\mathrm{p}}\otimes \mathcal I \left(\ket{\kappa}\bra{\kappa}_{\mathrm{TMSV}}\right)\right]
\end{equation}
is a linear combination of quadrature covariances.
From Eq.~(\ref{photonNumberToQuadrature}) and transformation~(\ref{HeisenbergPictureSqueezingOperation}),
we have
\begin{equation}
    W_{\mathrm{amp}}= \frac{\lambda+1}{g^2} \left[\mathds{1}-\frac{\lambda+1}{g^2} \hat{\bm{x}}_{\mathrm{A}'}^\top\hat{\bm{x}}_{\mathrm{A}'}
    +\frac{\sqrt{\lambda+1}}{g}\hat{\bm{x}}_{\mathrm{R}}^\top\bm{Z}\hat{\bm{x}}_{\mathrm{A}'}-\frac{1}{2} \hat{\bm{x}}_{\mathrm{R}}^\top\hat{\bm{x}}_{\mathrm{R}} +\frac{g^2-\lambda-1}{2g^2}\right].
\end{equation}
Combining~Eqs.~(\ref{TanhThetaPrime}) and~(\ref{referenceModeThermal}) yeilds
\begin{equation}\label{AmplifcationWitnessInQuadratureBasis}
\omega_{\mathrm{amp}}(\mathcal E_{\mathrm{p}})
=\frac{\lambda+1}{g^2}\left[\frac{(\lambda-4)g^2-\lambda^2-\lambda}{2\lambda g^2}-\frac{\lambda+1}{g^2} \left(\braket{\hat{q}_{\mathrm{A}'}^2}+\braket{\hat{p}_{\mathrm{A}'}^2}\right)+\frac{\sqrt{\lambda+1}}{g} \left(\braket{ \hat{q}_{\mathrm{A}'}\hat{q}_{\mathrm{R}}}-\braket{ \hat{p}_{\mathrm{A}'}\hat{p}_{\mathrm{R}}}\right)\right].
\end{equation}
Eq.~(\ref{AmplifcationWitnessInQuadratureBasis}) implies that the mean value of the average-fidelity witness can be estimated by sampling the covariances of the quadrature operators, as shown in Algorithm~\ref{alg:amplificationchannel}.

We have presented the verification protocols of two typical kinds of bosonic channels as examples of the general framework in section~\ref{sec:def}. 
Rather than estimating the average fidelity directly, both two verification protocols estimate the mean value of an average-fidelity witness, which ascertains an lower bound the average fidelity. 
The measurement of the average-fidelity witness requires only the preparation of two-mode squeezed vacuum states and the application of homodyne detections.
As the measurements on the reference modes can be applied immediately after the preparation of two-mode squeezed vacuum states, our verification protocols do not require any quantum memory to remain the entanglement between the channel-output modes and the reference modes.
The sample complexities of both quantum channels and two-mode squeezed vacuum state inputs in both two protocols are efficient with respect to all specification parameters of the target channels.

\section{\label{sec:dis}Discussion}
We have presented a general verification framework for an optimal quantum channel by unifying the favourable features of quantum-state verification~\cite{aolita2015reliable} and quantum-process benchmarking~\cite{bai2018test}. 
To develop our quantum-channel-verification framework, standard fidelity witness for quantum states has been generalized to an average fidelity witness for quantum channels per Definition~\ref{channelPerformanceWitness}. 
Rather than sampling a set of input states, our quantum-channel verification protocols require only one certain entangled input state and local measurements of an average-fidelity witness.
Our verification protocols satisfy both completeness and soundness conditions per Definition~\ref{def:quantum-channel-verification}, hence are reliable quantum-channel verification schemes.


We have presented the applications of our framework for the verification of two types of CPTP maps: multi-mode Gaussian unitary channels and single-mode amplification channels, both used widely in continuous-variable quantum computing and quantum communication. 
We devise average-fidelity witnesses for these two types of quantum channels in Theorems~\ref{theorem:GaussianUnitary} and Theorem~\ref{thm:amplificationchannel}, respectively, by truncating a thermal-state density operator in Lemma~\ref{lemma:GaussianFidelityWitness} and reformulating the witness in terms of quadrature operators. 
Sample complexity for verifying multi-mode Gaussian unitary channels scales polynomially with respect to number of modes~$m$, maximum squeezing~$\norm{\bm{S}}_{\infty}$ and phase-space displacement~$\norm{\bm{d}}$. On the other hand, sample complexity to verify single-mode amplification channels scales polynomially with respect to amplification gain~$g$. 
Sample complexities in both verification protocols are proportional to $\frac{1}{\epsilon^2 \ln\left(1/(1-\delta)\right)}$ due to classical sampling error.
Our measurement procedure comprises only local homodyne detections and is much simpler than the related work~\cite{bai2018test}, as neither online two-mode squeezing nor quantum memories are required.


\section{\label{sec:con}Conclusion}

We have presented experimentally feasible verification protocols for bosonic channels with polynomially scaling sample complexities. Different from quantum process tomography, our verification protocol's benchmark is average fidelity over an infinite set of gaussian-distributed coherent states. Our experimental setting uses only two-mode squeezed vacuum states and local homodyne detections, which are feasible using current technology. Our verification protocols are reliable in the sense that a  deceitful prover fails to cheat a prover and an honest prover typically passes the prover's test.

The essential step of our verification protocols is to measure an average-fidelity witness,
whose mean value can distinguish an optimal quantum channel from all other quantum channels, whose average fidelity is below a certain threshold. 
We apply our quantum-channel verification framework to verifying both multi-mode Gaussian unitary channels and single-mode amplification channels.
Owing to extensive usage of Gaussian unitary operations, like squeezing, in continuous-variable quantum information processing and the remarkable utilization of amplification channels in quantum communication~\cite{PhysRevA.86.012327,xiang2010heralded}, our verification protocols are important for testing components in continuous-variable quantum computing and quantum communication.

Our quantum-channel-verification framework can be applied to verify other types of quantum channels, for example, attenuation channels and optimal quantum cloning machines~\cite{cochrane2004optimal}. 
Furthermore, our approach can be extended to verify non-Gaussian cubic phase gates~\cite{PhysRevA.64.012310,weedbrook2012gaussian}, which is essensial for universal CV quantum computing, by estimating higher-order quadrature cumulants~\cite{liu2018client,farias2018average}.
Sample complexity, introduced here, can be 
further reduced by restricting the nature of the quantum channel and using statistical techniques, like importance sampling~\cite{gluza2018fidelity,farias2018average}. 
As this paper mainly focuses on CV quantum information, verification of linear optical devices for the significant application of BosonSampling, is not studied here, however, is an interesting direction to explore and could be quite related to our work here.
In the future, benchmark and verification protocols that does not rely on assuming independent and identical copies and are robust to SPAM errors will be important for continuous-variable quantum gates.

\section{Acknowledgments}

We thank Si-Hui Tan, Nana Liu and Yunlong Xiao for
their valuable discussions and acknowledge funding from NSERC.

\appendix

\section{Proof of Lemma~\ref{lemma:AverageFidelityObservable}}\label{app:observable}

The purification of thermal state~$\rho_{\mathrm{A}}=\rho_T(\frac{1}{\lambda})$ is a two-mode squeezed vacuum state 
\begin{equation}
\ket{\Psi}_{\mathrm{AR}}=\sqrt{\frac{\lambda}{1+\lambda}}\sum_{n=0}^\infty \left(\frac{1}{1+\lambda}\right)^{\frac{n}{2}}\ket{n}_{\mathrm{A}}\ket{n}_{\mathrm{R}}.
\end{equation}
The reduced states on~A and~R are
\begin{equation}\label{reducedstateAR}
    \rho_{\mathrm{A}}=\rho_{\mathrm{R}}=\frac{\lambda}{1+\lambda}\sum_{n=0}^\infty \left(\frac{1}{1+\lambda}\right)^n \ket{n}\bra{n}.
\end{equation}
Thus, \begin{equation}\label{partialisometry}
    T_{\mathrm{AR}}=\mathds{1}
\end{equation}
is an identity map on~$\mathscr{H}$.

Plugging Eqs.~(\ref{performanceOperatorAmplification}),~(\ref{reducedstateAR}) and~(\ref{partialisometry}) into Eq.~(\ref{benchmarkObservable}),
we obtain~\cite{bai2018test}
\begin{equation}\label{observableTobeMeasured}
    O_{\mathrm{A}'\mathrm{R}}=\int\frac{\mathrm{d}^2 \alpha}{\pi} \Big|\frac{g\alpha}{\sqrt{\lambda+1}}\Big\rangle\Big\langle\frac{g\alpha}{\sqrt{\lambda+1}}\Big|\otimes \ket{\bar{\alpha}}\bra{\bar{\alpha}}.
\end{equation}
If $g\le \sqrt{\lambda+1}$,
we have
\begin{equation}
     \forall \alpha\in \mathbb{C}, \;
     S_{\theta} \mathds{1}\otimes D\left(\frac{\bar{\alpha}}{\cosh\theta}\right)S_{\theta}^\dagger=D\left(\frac{g\alpha}{\sqrt{\lambda+1}}\right)\otimes D(\bar{\alpha}).
\end{equation}
Then~$O_{\mathrm{A}'\mathrm{R}}$~(\ref{observableTobeMeasured}) can be further simplified to
\begin{align}\nonumber
    O_{\mathrm{A}'\mathrm{R}}=&\int \frac{\mathrm{d}^2 \alpha}{\pi} S_{\theta} \mathds{1}\otimes D\left(\frac{\bar{\alpha}}{\cosh\theta}\right)S_{\theta}^\dagger \ket{0}\bra{0}\otimes \ket{0}\bra{0} 
S_{\theta} \mathds{1}\otimes D\left(\frac{\bar{\alpha}}{\cosh\theta}\right)^\dagger S_{\theta}^\dagger \\ \nonumber
=&\cosh^2\theta \int \frac{\mathrm{d}^2 \alpha}{\pi} S_\theta \left(\mathds{1}\otimes D(\alpha)\right) S_\theta^\dagger \ket{0}\bra{0}\otimes \ket{0}\bra{0} 
S_\theta \left(\mathds{1}\otimes D(\alpha)^\dagger \right) S_\theta^\dagger \\ \label{observable1design}
=& S_\theta G_\theta \otimes \mathds{1} S_\theta^\dagger.
\end{align}
In Eq.~(\ref{observable1design}), we use the fact that the Heisenberg-Weyl group forms a unitary $1$-design~\cite{blume2014curious,zhuang2019scrambling}; i.e.,
\begin{equation}\label{CV1design}
    \int\frac{\mathrm{d}^2 \alpha}{\pi} D(\alpha) \rho D(\alpha)^\dagger =\mathds{1},
\end{equation}
for any single-mode density operator~$\rho$.

If $g\ge \sqrt{\lambda+1}$,
\begin{equation}
    \forall \alpha\in \mathbb{C}, \;
    S_{\theta'} \left( D\left(\frac{\alpha}{\sinh\theta'}\right) \otimes \mathds{1}\right) S_{\theta'}^\dagger=D\left(\frac{g\alpha}{\sqrt{\lambda+1}}\right)\otimes D(\bar{\alpha}),
\end{equation}
for $\theta'=\arctanh\frac{\sqrt{\lambda+1}}{g}$.
$O_{\mathrm{A}'\mathrm{R}}$ in Eq.~(\ref{observableTobeMeasured}) can be simplified to
\begin{align}\nonumber
O=&\int \frac{\mathrm{d}^2 \alpha}{\pi} S_{\theta'} \left( D\left(\frac{\alpha}{\sinh{\theta'}}\right) \otimes \mathds{1}\right) S_{\theta'}^\dagger \ket{0}\bra{0}\otimes \ket{0}\bra{0} 
S_{\theta'} \left( D\left(\frac{\alpha}{\sinh{\theta'}}\right)^\dagger \otimes \mathds{1} \right) S_{\theta'}^\dagger \\ \nonumber
=&\sinh^2{\theta'} \int \frac{\mathrm{d}^2 \alpha}{\pi} S_{\theta'} \left( D(\alpha)\otimes  \mathds{1}\right) S_{\theta'}^\dagger \ket{0}\bra{0}\otimes \ket{0}\bra{0} 
S_{\theta'} \left(D(\alpha)^\dagger \otimes  \mathds{1}\right) S_{\theta'}^\dagger \\\label{observable1designagain}
=&\tanh^2{\theta'} S_{\theta'}  \mathds{1}\otimes G_{\theta'} S_{\theta'}^\dagger,
\end{align}
where we use Eq.~(\ref{CV1design}) again to obtain Eq.~(\ref{observable1designagain}).
Thus, we have proved Lemma~\ref{lemma:AverageFidelityObservable}.

\section{Proof of Lemma~\ref{lemma:GaussianFidelityWitness}}\label{proofLemma}
\begin{proof}
We first prove that~$G_{\theta}\ge\mathds{1}-\frac{\hat{n}}{\bar{n}_T+1}$.
This can be seen by
\begin{align}\nonumber
\mathds{1}-\frac{\hat{n}}{\bar{n}_T+1}=&\sum_{n=0}^\infty \left( 1-\frac{n}{\bar{n}_T+1}\right)\ket{n}\bra{n} \\\nonumber
=&\sum_{n=0}^\infty \frac{\bar{n}_T+1-n}{\bar{n}_T+1}\ket{n}\bra{n} \\ \label{lowerbound}
=&\sum_{n=0}^\infty (1-n\sech^{2}\theta) \ket{n}\bra{n}.
\end{align}
From the binomial inequality,
\begin{equation}
1-n\sech^{2}\theta\le \left(1-\sech^2\theta\right)^n=\tanh^{2n} \theta.
\end{equation}
Combining Eqs.~(\ref{GthetaOperator}) and~(\ref{lowerbound}), we have
\begin{equation}
    G_{\theta}\ge\mathds{1}-\frac{\hat{n}}{\bar{n}_T+1}.
\end{equation}
Next
we use this result to prove the lemma by induction.
Suppose 
\begin{equation}
G_\theta^{\otimes (m-1)} \ge \mathds{1}-\frac{\sum_{i=1}^{m-1}\hat{n}_i}{\bar{n}_T+1},
\end{equation} then
\begin{equation}
G_\theta^{\otimes m} \ge \left(\mathds{1}-\frac{\sum_{i=1}^{m-1}\hat{n}_i}{\bar{n}_T+1}\right)\left(\mathds{1}-\frac{\hat{n}_n}{\bar{n}_T+1}\right)
\ge \mathds{1}-\frac{\sum_{i=1}^m\hat{n}_i}{\bar{n}_T+1}.
\end{equation}
Thus, we have proved Lemma~\ref{lemma:GaussianFidelityWitness}.
\end{proof}

\section{Sample complexity for verification of Gaussian unitary channels}\label{sampleComplexityGaussianUnitary}

We denote the estimation errors as
 \begin{align}
     \bm{\epsilon}\coloneqq&\bm{\gamma}-\bm{\gamma}^*,\\
     \bm{E}_1\coloneqq&\bm{\Gamma}_1-\bm{\Gamma}_1^*,\\
     \bm{E}_{2}\coloneqq&\bm{\Gamma}_{2}-\bm{\Gamma}_{2}^*.
 \end{align} 
The distance between $w$ and experimental value $w^*$ can be bounded
\begin{align}
\left|\omega_{U_{\bm{S}, \bm{d}}}(\mathcal E_{\mathrm{p}}) -\omega_{U_{\bm{S}, \bm{d}}}(\mathcal E_{\mathrm{p}})^*\right|\le 
&\frac{1}{2}\left| \tr \left[ \bm{S}^{-\mathrm{T}}\bm{S}^{-1} \left( \bm{E}_1 -2 \bm{\epsilon} \bm{d}^\top
\right)\right] \right| +\frac{1}{\sqrt{\lambda+1}}  \left|\tr \left(\bm{Z}^{\oplus m} \bm{S}^{-1} \bm{E}_{2}\right)\right|\\ \label{secondInequalityinErrorUnitary}
\le & \frac{1}{2} \norm{\bm{S}^{-\mathrm{T}}\bm{S}^{-1}}_{\infty}\norm{\bm{E}_1 -2 \bm{\epsilon} \bm{d}^\top}_1 +
\frac{1}{\sqrt{\lambda+1}}\norm{\bm{Z}^{\oplus m} \bm{S}^{-1}}_\infty\norm{\bm{E}_{2}}_1 \\\label{thirdInequalityinErrorUnitary}
\le & \frac{1}{2} \norm{\bm{S}^{-\mathrm{T}}\bm{S}^{-1}}_{\infty}(\norm{\bm{E}_1}_1 +2 \norm{\bm{\epsilon}}_1 \norm{\bm{d}}_1) +
\frac{1}{\sqrt{\lambda+1}} \norm{\bm{S}^{-1}}_{\infty}\norm{\bm{E}_{2}}_1,
\end{align}
where we use 
\begin{equation}
    \left|\tr(\bm{A}\bm{B})\right|\le \norm{\bm{A}}_{\infty}\norm{\bm{B}}_1
\end{equation} in~(\ref{secondInequalityinErrorUnitary}), and 
\begin{equation}
    \norm{\bm{A}\bm{B}}_1\le \norm{\bm{A}}_1\norm{\bm{B}}_1
\end{equation} in~(\ref{thirdInequalityinErrorUnitary})
for any matrices~$\bm{A}$ and~$\bm{B}$.

From the singular value decomposition of the symplectic matrix~$\bm{S}$, we obtain 
\begin{equation}
    \norm{\bm{S}^{-1}}_{\infty}=\norm{\bm{S}}_{\infty}
\end{equation} and 
\begin{equation}
    \norm{\bm{S}^{-\mathrm{T}}\bm{S}^{-1}}_{\infty}=\norm{\bm{S}}_{\infty}^2.
\end{equation}
Plugging the inequalities
\begin{align}
    \norm{\bm{E}_1}_1\le& 2 m\norm{\bm{E}_1}_{\mathrm{max}}, \\
    \norm{\bm{E}_{2}}_1\le& 2m \norm{\bm{E}_{2}}_{\mathrm{max}},\\
    \norm{\bm{d}}_1\le& \sqrt{2m}\norm{\bm{d}},\\
    \norm{\bm{\epsilon}}_1\le& 2m\norm{\bm{\epsilon}}_\infty,
\end{align}
into Eq.~(\ref{thirdInequalityinErrorUnitary}), we have
\begin{equation}\label{estimation-error-bound}
\left|\omega_{U_{\bm{S}, \bm{d}}}(\mathcal E_{\mathrm{p}}) -\omega_{U_{\bm{S}, \bm{d}}}(\mathcal E_{\mathrm{p}})^*\right|\le  (2m)^{\frac{3}{2}} \norm{\bm{S}}_{\infty}^2\norm{\bm{\epsilon}}_\infty \norm{\bm{d}} +m \norm{\bm{S}}_{\infty}^2\norm{\bm{E}_1}_{\mathrm{max}} +
\frac{2 m\norm{\bm{S}}_{\infty}}{\sqrt{\lambda+1}} \norm{\bm{E}_{2}}_{\mathrm{max}}.
\end{equation}
To guarantee that
\begin{equation}
    P\left(\left|\omega_{U_{\bm{S}, \bm{d}}}(\mathcal E_{\mathrm{p}}) -\omega_{U_{\bm{S}, \bm{d}}}(\mathcal E_{\mathrm{p}})^*\right|\le \epsilon\right)\ge 1-\delta,
\end{equation} where $P(\cdot)$ denotes the probability of an event, we suppose each term on the right-hand side of~(\ref{estimation-error-bound}) is less than~$\frac{\epsilon}{3}$ with probability no less than~$(1-\delta)^{\frac{1}{3}}$.
To determine sample complexity, we use the following lemma~\cite{aolita2015reliable}.
\begin{lemma}\label{lemma:samplecomplexity}
Suppose $O_1, O_2, \dots, O_l$ are observables on state~$\rho$ with mean values 
\begin{equation}
    m_j\coloneqq\tr(O_j \rho),
\end{equation}
and variances bound by $\sigma>0$; i.e.,
\begin{equation}
    \forall j,\; \tr(O_j^2 \rho)-m_j^2\le \sigma.
\end{equation}
For each~$j$, $\chi_i^{O_j}$ denotes the $i$th measurement outcome of~$O_j$ on~$\rho$, and then the finite sample mean over $c$ measurements of $O_j$ is
\begin{equation}
    m_j^*=\frac{1}{c}\sum_{i=1}^c \chi_i^{O_j}.
\end{equation}
For any $\epsilon>0$, $0<\delta\le\frac{1}{2}$, to make
\begin{equation}
    P\left(\forall j, |m_j^*-m_j|\le \epsilon \right)\ge 1-\delta,
\end{equation}
the number of measurements should satisfy that
\begin{equation}
    c\ge\frac{\sigma^2(l+1)}{\epsilon^2\ln\left(1/(1-\delta)\right)}.
\end{equation}
\end{lemma}
\noindent From this lemma, we know that, to make 
\begin{equation}
P\left((2m)^{\frac{3}{2}} \norm{\bm{S}}_{\infty}^2\norm{\bm{\epsilon}}_\infty \norm{\bm{d}} \le \frac{\varepsilon}{3}\right)\ge (1-\delta)^\frac{1}{3},
\end{equation}
the verifier applies $c_3$~(\ref{GaussianSampleComplexity1})
measurements on each $\hat{\bm{x}}_l^{\mathrm{A}'} (1\le l\le 2m)$, respectively, to estimate~$\bm{\gamma}$.
Similarly, to make 
\begin{equation}
P\left(m \norm{\bm{S}}_{\infty}^2 \norm{\bm{E}_1}_{\mathrm{max}} \le \frac{\varepsilon}{3} \right)\ge (1-\delta)^\frac{1}{3},
\end{equation}
and 
\begin{equation}
P\left(\frac{2m\norm{\bm{S}}_{\infty}}{\sqrt{\lambda+1}} \norm{\bm{E}_{2}}_{\mathrm{max}} \le \frac{\varepsilon}{3} \right)\ge (1-\delta)^\frac{1}{3},
\end{equation}
the verifier applies $c_4$~(\ref{GaussianSampleComplexity2})
 measurements on each $\frac{1}{2}\left(\hat{\bm{x}}^{\mathrm{A}'}_u \hat{\bm{x}}^{\mathrm{A}'}_v+\hat{\bm{x}}^{\mathrm{A}'}_v\hat{\bm{x}}^{\mathrm{A}'}_u\right)$,  and~$c_5$~(\ref{GaussianSampleComplexity3})
 measurements on each $\hat{\bm{x}}_u^{\mathrm{A}'}\hat{\bm{x}}_v^{\mathrm{R}}$, where
$1\le u, v\le 2m$.

\section{Sample complexity for verification of amplification channels}

We denote the estimation errors of
$\braket{ \hat{q}_{\mathrm{A}'}^2}$, $\braket{\hat{p}_{\mathrm{A}'}^2}$, $\braket{ \hat{q}_{\mathrm{A}'}\hat{q}_{\mathrm{R}}}$ and $\braket{ \hat{p}_{\mathrm{A}'}\hat{p}_{\mathrm{R}}}$ as~$E_{1}$, $E_{2}$, $E_{11}$ and $E_{22}$.
The estimation error between~$\omega(\mathcal E_{\mathrm{p}})$ and $\omega(\mathcal E_{\mathrm{p}})^*$ is bounded by
\begin{equation}\label{errorboundamplification}
\left|\omega(\mathcal E_{\mathrm{p}})-\omega(\mathcal E_{\mathrm{p}})^*\right|\le\frac{g^2}{\lambda+1}  \mathrm{max}\{ \norm{E_{1}}, \norm{E_{2}}\}+
2\left(\frac{g}{\sqrt{\lambda+1}}\right)^3 \mathrm{max}\{\norm{E_{11}}, \norm{E_{22}}\} .
\end{equation}
To make 
\begin{equation}
    P\left(\left|\omega(\mathcal E_{\mathrm{p}})-\omega(\mathcal E_{\mathrm{p}})^*\right|\le \epsilon\right)\ge 1-\delta,
\end{equation}
we suppose each term at the right-hand side of~(\ref{errorboundamplification}) is less than~$\frac{\epsilon}{2}$ with probability no less than~$\sqrt{1-\delta}$.
From Lemma~\ref{lemma:samplecomplexity}, we know that, to make 
\begin{equation}
    P\left(\frac{g^2}{\lambda+1}  \mathrm{max}\{ \norm{E_{1}}, \norm{E_{2}}\} \le \frac{\epsilon}{2}\right)\ge \sqrt{1-\delta},
\end{equation}
and
\begin{equation}
    P\left(2\left(\frac{g}{\sqrt{\lambda+1}}\right)^3 \mathrm{max}\{\norm{E_{11}}, \norm{E_{22}}\} \le \frac{\epsilon}{2}\right)\ge \sqrt{1-\delta},
\end{equation}
the verifier needs~$c_6$~(\ref{amplificationSampleComplexity1})
measurements on~$\hat{q}_{\mathrm{A}'}^2$ and~$\hat{p}_{\mathrm{A}'}^2$, respectively,
and~$c_7$~(\ref{amplificationSampleComplexity2}) measurements on $\hat{q}_{\mathrm{A}'}\hat{q}_{\mathrm{R}}$ and $\hat{p}_{\mathrm{A}'}\hat{p}_{\mathrm{R}}$, respectively.
 
\section{Measurements by local homodyne detections}\label{measurementscheme}

For each $1\le l\le 2m$, mean value of $\hat{x}_l$ can be sampled by a local homodyne detection on either position or momentum basis.
Sampling~$2m$ quadrature mean values require two local homodyne settings: one is measuring position on all~$m$ modes of~A', the other is measuring momentum on all~$m$ modes of~A'. 
For each $1\le u, v\le 2m$,  mean value of $\hat{\bm{x}}^{\mathrm{A}'}_u \hat{\bm{x}}^{\mathrm{R}}_v$ can be sampled by performing local homodyne detections regarding~$\hat{\bm{x}}^{\mathrm{A}'}_u$ and $\hat{\bm{x}}^{\mathrm{R}}_v$, respectively, and then multiplying two measurement outcomes. 
Sampling mean values of $\hat{\bm{x}}^{\mathrm{A}'}_u \hat{\bm{x}}^{\mathrm{R}}_v$ require two additional homodyne settings: one is measuring position on all~$m$ modes of~R; the other is measuring momentum on all~$m$ modes of~R.

For each $1\le v< u\le 2m$, such that $(u, v)\neq(2j, 2j-1)$ for $1\le j\le m$, sampling mean value of $\hat{\bm{x}}^{\mathrm{A}'}_u \hat{\bm{x}}^{\mathrm{A}'}_v$ can be accomplished by applying local homodyne detections regarding~$\hat{\bm{x}}^{\mathrm{A}'}_u$ and~$\hat{\bm{x}}^{\mathrm{A}'}_v$, respectively, and then multiplying measurement outcomes. These measurements need the combination of position measurement at one mode and momentum measurement at another mode. Hence, at least $m$ more local homodyne settings are required: each one setting measures position at one distinct mode and momenta at all other modes.

For each $1\le u\le 2m$, sampling mean value of $\left(\hat{\bm{x}}_u^{\mathrm{A}'}\right)^2$ can be accomplished by performing homodyne detection with respect to $\hat{\bm{x}}^{\mathrm{A}'}_u$
and squaring the measurement outcomes. These homodyne settings are same as the settings for sampling mean values of~$\hat{\bm{x}}_u^{\mathrm{A}'}$. 
When $(u, v)=(2j, 2j-1)$, $\frac{1}{2}\left(\hat{\bm{x}}^{\mathrm{A}'}_u \hat{\bm{x}}^{\mathrm{A}'}_v+\hat{\bm{x}}^{\mathrm{A}'}_v\hat{\bm{x}}^{\mathrm{A}'}_u\right)$ 
is 
\begin{equation}\label{positionMomentumProduct}
\frac{1}{2}\left(\hat{q}_j^{\mathrm{A}'}\hat{p}_j^{\mathrm{A}'}+\hat{p}_j^{\mathrm{A}'}\hat{q}_j^{\mathrm{A}'}\right).
\end{equation}
To sample mean value of observable~(\ref{positionMomentumProduct}), one can sample mean value of
\begin{equation}\label{linearCombinationofPositionMomentum}
     \frac{1}{\sqrt{2}}(\hat{q}_j^{\mathrm{A}'}+\hat{p}_j^{\mathrm{A}'}),
\end{equation}
by noting that
\begin{equation}
\frac{1}{2}\left(\hat{q}\hat{p}+\hat{p}\hat{q}\right)=\frac{1}{2}(\hat{q}+\hat{p})^2-\frac{1}{2}\hat{q}^2-\frac{1}{2}\hat{p}^2,
\end{equation}
and that mean value of $\left(\hat{q}_j^{\mathrm{A}'}\right)^2$ and $\left(\hat{p}_j^{\mathrm{A}'}\right)^2$ have been sampled by the approach we explained above.  
Sampling mean value of observable~(\ref{linearCombinationofPositionMomentum}), for each~$j$,
can be accomplished by one additional measurement setting that is to perform homodyne detection at each mode of~A' in a $45$-degree rotated basis. 
Thus, all measurements in Algorithm~\ref{alg:GaussianUnitaryOperation} can be accomplished by $m+5$ local homodyne settings.
 
\bibliography{reference.bib}
\end{document}